\documentclass[onecolumn,epjc3,envcountsect]{svjour3} 
\smartqed  % Flush right qed marks, e.g. at end of proof

\journalname{Eur. Phys. J. C}

\usepackage{graphicx}
\usepackage{subcaption}
\usepackage{amsmath,amssymb}
\usepackage{esint}
\usepackage{gensymb}
\usepackage{multirow}
\usepackage{cancel}
\usepackage{xfrac}
\usepackage{flushend} % Highly recommended by Springer to balance the two columns on the final page

\usepackage{tikz}
\usepackage{pgfplots}
\pgfplotsset{compat=1.18}
\usetikzlibrary{positioning,patterns,calc,decorations.markings}

\usepackage{tikz-feynman}
\tikzfeynmanset{compat=1.1.0, warn luatex=false}

\makeatletter
\def\tikzfeynman@luatex@required@path{
  \PackageWarning{tikz-feynman}{The key you tried to use '\pgfkeyscurrentpath/\pgfkeyscurrentname' requires LuaTeX. It will be ignored.}
}
\def\tikzfeynman@luatex@required@key{
  \PackageWarning{tikz-feynman}{The key you tried to use '\pgfkeyscurrentpath' requires LuaTeX. It will be ignored.}
}
\makeatother

\usepackage[dvipsnames]{xcolor}
\usepackage[colorlinks=true, linkcolor=blue, citecolor=blue, urlcolor=blue]{hyperref}
\usepackage{cleveref}

\usepackage[section]{placeins}
\usepackage[numbers,sort&compress]{natbib}

\newcommand{\kperp}{k_{\perp}}
\newcommand{\qperp}{q_{\perp}}
\newcommand{\qperpone}{q_{\perp,1}}
\newcommand{\qperptwo}{q_{\perp,2}}
\newcommand{\pperp}{p_{\perp}}
\newcommand{\qperpvec}{\boldsymbol{q}}
\newcommand{\pperpvec}{\boldsymbol{p}}
\newcommand{\bb}{\boldsymbol{b}}
\newcommand{\muperp}{\mu_{\perp}}
\newcommand{\pvec}{\vec{p}}

\begin{document}

\title{Momentum Broadening in the Opacity Expansion: All-Path-Length Corrections and Improved Regge Kinematics}

\author{Dario van den Berg\thanksref{e1,addr1,addr2,addr3}
        \and
        Isobel Kolb\'{e}\thanksref{e2,addr1,addr2,addr3} %etc.
}


\thankstext{e1}{e-mail: dario.vandenberg@gmail.com}
\thankstext{e2}{e-mail: isobel.kolbe@wits.ac.za}

\institute{%
  \parindent0pt\relax
   \label{addr1} School of Physics, University of the Witwatersrand, Johannesburg, 2000, South Africa \and
   \label{addr2} Mandelstam Institute for Theoretical Physics (MITP), University of the Witwatersrand, Johannesburg, 2000, South Africa \and
   \label{addr3} National Institute for Theoretical and Computational Sciences (NITheCS), Stellenbosch, 7600, South Africa
}

\date{Received: date / Accepted: date}

\maketitle

\begin{abstract}
We present a detailed study of momentum broadening and the jet transport 
coefficient $\hat{q}$ for high-energy partons traversing the Quark-Gluon 
Plasma (QGP), extending the Gyulassy-Levai-Vitev (GLV) formalism to include 
both all-path-length (APL) and sub-Regge kinematical corrections. Traditional 
GLV calculations rely on the large separation distance and large formation 
time approximations, which are valid for large systems but whose applicability 
to small systems, such as $pp$ and $p/d$A collisions, may fail. We derive 
analytic expressions for the momentum broadening distributions and $\hat{q}$ 
to first order in the opacity expansion, and perform a detailed numerical 
investigation to quantify their impact. The APL correction suppresses 
momentum broadening at low $p_{\perp}$, with a correction scaling as 
$\propto 1/(L\, p_{\perp})$ that dominates for small systems, while 
converging to the standard GLV result at large $L$. The sub-Regge kinematical 
correction enhances momentum broadening at high $p_{\perp}$, becoming 
significant when the transverse momentum transfer approaches the magnitude 
of the parton's large light-cone momentum component, and vanishing in the 
Regge limit where this ratio is small. When both corrections are 
combined, the sub-Regge kinematical correction partially mitigates the 
suppression induced by the APL term; in the case of $\hat{q}$, this 
mitigation is essentially complete, with $\hat{q}_{(\mathrm{APL+SUB})}$ 
found to coincide closely with the standard GLV result. These findings demonstrate that sub-Regge kinematical corrections can resolve 
the long-standing problem of large negative energy-loss contributions at 
high energies identified in earlier studies.
\end{abstract}

\section{Introduction}

    The study of quark--gluon--plasma (QGP) formation has traditionally focused on relativistic heavy-ion collisions, where large systems of deconfined matter are produced under extreme energy densities \cite{ALICE:2022wpn}. The QGP is a strongly interacting phase of QCD in which quarks and gluons become deconfined over extended volumes, forming a nearly perfect fluid with extremely low shear viscosity over entropy density $\eta/s$ \cite{Heinz:2013th,Busza:2018rrf}. Its properties are encoded in transport coefficients and response functions that can be probed by hard partons produced in the early stages of the collision.
    
    High-energy quarks and gluons traversing the QGP are predicted to undergo multiple scatterings, leading to jet quenching, a suppression and modification of high-$p_{T}$ observables relative to proton--proton expectations \cite{Majumder:2010qh,Qin:2015srf}. Among others, jet quenching is observed through reduced yields of high-$p_{T}$ hadrons and jets \cite{ALICE:2015mdb,ATLAS:2022iyq,CMS:2011iwn,Spousta:2013aaa,STAR:2017hhs} and medium-induced modifications to jet substructure \cite{ALICE:2024jtb,ATLAS:2025svn,CMS:2025dnx,STAR:2023pal}. A key component of jet quenching is momentum broadening, the stochastic transverse momentum accumulation of a propagating parton due to repeated soft scatterings with medium constituents \cite{Baier:1996kr,Casalderrey-Solana:2007ahi}.% The jet transport coefficient $\hat{q}$, defined as the average transverse momentum squared acquired per unit path length, characterizes this broadening and serves as a fundamental measure of the QGP's microscopic structure \cite{Baier:1996kr,Casalderrey-Solana:2007ahi}.\\
    
    Perturbative Quantum Chromodynamics (pQCD)--based energy loss models \cite{Wang:2025lct} such as GLV \cite{Gyulassy:1999ig,Gyulassy:1999bp,Gyulassy:1999zd,Gyulassy:2000fs,Gyulassy:2000er,Vitev:2002vr}, BDMPS-Z \cite{Baier:1994bd,Baier:1996kr,Baier:1996sk,Baier:1998kq}, and ASW \cite{Armesto:2003jh}  have been extremely successful in quantitatively describing the observed jet quenching patterns in nucleus--nucleus (AA) collisions at RHIC and the LHC \cite{Connors:2017ptx,Mehtar-Tani:2013pia}. These frameworks incorporate both collisional and radiative energy loss \cite{Cao:2020wlm} and have achieved remarkable consistency across diverse observables, such as the nuclear modification factor $R_{AA}$ \cite{Qin:2007rn} and jet shape broadening \cite{Chang:2019sae}. Their success has made pQCD energy-loss formalisms central tools for extracting QGP transport coefficients, and for understanding the behavior of QCD matter at extreme temperatures and densities.
    
    However, recent experimental results have revealed QGP-like signatures even in small collision systems such as high-multiplicity proton--proton (pp) and proton--lead (pPb) collisions. Observables including collective multi-particle correlations \cite{Nagle:2018nvi}, strangeness enhancement \cite{ALICE:2024zxp}, and quarkonium suppression \cite{ALICE:2015kgk}, previously considered exclusive to large systems, have also been reported in much smaller collision systems. These findings raise fundamental questions about the minimal conditions necessary for QGP formation, the role of initial-state correlations or saturation physics, and whether a short-lived, deconfined medium can emerge in systems with limited size and lifetime \cite{Nagle:2018nvi}. For reviews of extensive results, see \cite{Wang:2025lct}.
    
    The first oxygen-oxygen, proton-oxygen, and neon-neon collision runs at the LHC highlight the growing need to understand energy loss, transverse momentum broadening, and medium formation in intermediate-sized systems. These collisions provide unprecedented opportunities to probe the onset of QGP-like behavior with controlled variations in system size, multiplicity, and nuclear geometry. In particular, they allow tests of whether jet quenching, collectivity, and strangeness enhancement persist as the number of participants decreases, thereby offering key constraints on initial-state modeling, small-system hydrodynamics, and the possible emergence of a short-lived, strongly interacting medium \cite{Brewer:2021kiv,Casalderrey-Solana:2016jvj}. The $\text{O}^{16}\text{O}^{16}$ and $\text{Ne}^{20}\text{Ne}^{20}$ datasets thus play a crucial role in determining the applicability limits of pQCD energy-loss frameworks and in distinguishing genuine medium-induced phenomena from alternative explanations such as color reconnection or gluon saturation dynamics.
    
    Recent theoretical developments further motivate the possibility that $\text{O}^{16}\text{O}^{16}$ and $\text{Ne}^{20}\text{Ne}^{20}$ collisions may already lie within a regime where QGP-like dynamics become relevant \cite{Nijs:2021clz,Nijs:2020roc,Beattie:2022ojg}. The analysis in \cite{vanderSchee:2025hoe} further demonstrates that a single, smoothly varying medium parameter can describe nuclear modification patterns from pp and pA through $\text{O}^{16}\text{O}^{16}$ and up to PbPb. Such continuity strongly suggests that even intermediate-sized systems may achieve densities sufficient for partial color deconfinement and multiple scattering, supporting the interpretation that QGP-like matter can form below the traditionally assumed system-size threshold. 
    
    Further support for the relevance of medium-induced interactions in these systems comes from the global analysis presented in \cite{Faraday:2025pto,Faraday:2024qtl}, which highlights the essential role of transverse momentum broadening and elastic energy loss in describing suppression patterns across all collision systems. The extracted fits require nonzero elastic contributions even in the smallest systems, indicating nontrivial momentum broadening. In particular, the suppression exhibits a maximum around $p_T \sim$7 GeV, a transitional region where the probe is neither fully hydrodynamic nor asymptotically hard, and where purely Regge approximations are known to lose accuracy \cite{vanderSchee:2025hoe}. SUCH behavior naturally points to sensitivity to path-length–dependent scattering and sub-Regge kinematical corrections. The conclusions are consistent with recent theoretical studies emphasizing the importance of transverse momentum broadening, elastic scattering, and sub-Regge kinematical corrections in small and intermediate collision systems \cite{Pablos:2025cli}. A detailed theoretical treatment of momentum broadening and elastic scattering therefore becomes essential, as such effects may offer some of the earliest and most sensitive indicators of medium formation in new LHC and RHIC collision systems.
    
    Conventional jet quenching models, such as the Gyulassy--Levai--Vitev (GLV) formalism \cite{Gyulassy:1999ig,Gyulassy:1999bp,Gyulassy:1999zd,Gyulassy:2000fs,Gyulassy:2000er,Vitev:2002vr}, rely on several approximations that are valid in large systems but may break down in smaller collision systems. In the present work we will focus on two in particular:

        \begin{itemize}
            \item \textbf{Large separation distance:} In the GLV approach, the distance $\Delta z = z_1 - z_0$ between the hard parton production point $z_0$ and the first scattering center $z_1$ satisfies $\Delta z \gg \lambda_{mfp} \gg 1/\mu$, where $\lambda_{mfp}$ is the parton mean free path and $\mu$ the Debye screening mass. 
            While it is crucial that $\lambda_{mfp}\gg 1/\mu$, small colliding systems necessarily select for events in which $\Delta z \sim 1/\mu$.
            \item \textbf{Large formation time:} When including radiation, GLV  assumes that the gluon formation time $\tau \gg 1/\mu $, where $\tau \sim \frac{x P^+}{\kperp^2} $,  where $P^+$ is the initial parton momentum, $x$ is the momentum fraction carried by the radiated gluon, and $\kperp$ is the magnitude of the radiated gluon's  transverse momentum. %  %the transverse momentum transferred between the scattering centre and leading parton, an appropriate broadening analogue of the large formation time approximation would be \ik{fix this too..}$\omega \sim q_\perp^2 / P^+$.
            %In this form it becomes clear that the large formation time approximation is really a restatement of the high-energy, or Regge approximation.
            % \item \textbf{Regge approximation:} Standard calculations assume the hard parton follows a straight-line, lightlike trajectory, neglecting transverse deflection and finite longitudinal momentum transfer. While accurate for very high energies, this approximation may fail for the moderate-$p_T$ partons typical of small systems.
        \end{itemize}
    
    While broadening has been computed in the GLV formalism \cite{Gyulassy:1999ig,Gyulassy:1999bp,Gyulassy:1999zd,Gyulassy:2000fs,Gyulassy:2000er,Vitev:2002vr,Gyulassy:2002yv,Qiu:2003pm}, even including a flowing medium \cite{Sadofyev:2021ohn}, it is not straightforward to relax the above approximations. 
    The first attempt to correct for the large system approximation \cite{Kolbe:2015rvk} 
    revealed that, while the large formation time approximation played only a small role in the usual approximation scheme, when relaxing the large separation distance approximation, the large formation time approximation meant that the majority of diagrams did not have an all-path length contribution.
    That is to say that, due to the large formation time approximation, almost all path-length corrections remained suppressed.
    One of the major results of
    \cite{Kolbe:2015suq}
    was that the all path-length correction is negative and dominates at large energies, suggesting that dampening corrections are missing.
    
    Relaxing the large formation time approximation along with the large separation distance approximation in the full radative energy-loss case is an enormous undertaking.
    As such, in this work we will relax both approximations for the simpler broadening case.
    One may additionally argue that, since broadening (and quenching) effects are more pronounced at lower partonic energies \cite{Li:2023jeh}, and the effect is expected to be small in small systems due to the shorter pathlength \cite{Faraday:2025pto,Faraday:2024qtl}, one may consider partonic energies that are not infinitely higher than the momentum scale of the medium in order to detect quenching effects.
    However, this violates the broadening analogue of the large formation time approximation which requires $P^+ \gg \qperp$ (see \cite{Gyulassy:1999ig,Gyulassy:1999bp,Gyulassy:1999zd,Gyulassy:2000fs,Gyulassy:2000er,Vitev:2002vr,Gyulassy:2002yv,Qiu:2003pm}), further motivating the present work in which sub-Regge kinematical corrections are computed.
    
    In this work we will refer to the following corrections to the standard GLV calculation:
    % To address these limitations, we introduce two complementary corrections: 
        \begin{enumerate}
        \item \emph{All-path-length (APL) corrections}, which relax the large separation distance approximation by accounting for contributions from all possible scattering distances.
        \item \emph{Sub-Regge corrections}, which retain $1/P^+$ correction terms to the Regge kinematics, analogous to capturing the finite formation time effects in the radiative case.
        \end{enumerate}
    
    % These improvements provide a more accurate framework for studying high-$p_T$ parton propagation in small QGP systems and allow systematic comparisons with conventional GLV predictions. This work thus contributes to the ongoing effort to understand the onset of collectivity and partonic energy loss in the smallest strongly interacting systems accessible at RHIC and the LHC. 

We present the APL correction computed under both standard Regge and improved (Sub-Regge) kinematics. The results demonstrate that adopting the improved kinematics leads to a moderation of the APL correction, suggesting that the full radiative APL result may be similarly moderated upon the inclusion of Sub-Regge kinematical corrections (see \cref{fig:vsL} and \cref{fig:vsPpplot}), a treatment which lies beyond the scope of the present work.
    
    Relaxing the above approximations raises several technical questions.
    For discussions around the APL corrections, we refer the reader to \cite{Kolbe:2015rvk}.
    The sub-Regge contributions, which constitute the main novelty of the present work, appear to be closely related to the question of sub-eikonal corrections.
    However, we stress that the sub-Regge kinematical corrections (at the level of the eikonal diagrams considered here) are of a lower order than sub-eikonal corrections to the Wilson line resummations that have been studied extensively elsewhere \cite{Altinoluk:2020oyd,Altinoluk:2014oxa,Chirilli:2018kkw,Altinoluk:2015gia,Laenen:2010uz,Casalderrey-Solana:2007knd,Chirilli:2021,Altinoluk:2022,Sadofyev:2022,Kovchegov:2016,Kovchegov:2019a,Kovchegov:2019b,Kovchegov:2016b}.    
    Instead of truly considering sub-eikonal corrections to the Wilson line resummation, which would introduce new Feynman structures, we consider, in much the same spirit as \cite{Sadofyev:2021ohn}, corrections at the kinematical level that remain consistent within the eikonal approximation of the Wilson line (see Appendix A of \cite{Sadofyev:2021ohn} and related discussions).
    To see explicitly that the present kinematical corrections at the level of eikonal diagrams do not constitute sub-eikonal corrections, we show in \cref{crossed} that the first sub-eikonal diagram to contribute, the crossed diagram, contributes at a lower order than that considered here.

    The kinematical correction considered here enters at $\mathcal{O}(1/P^+)$ in the 
    amplitude $\mathcal{M}_1$, and consequently at $\mathcal{O}(1/P^{+2})$ in the interference 
    contribution $\text{Re}(\mathcal{M}_2\mathcal{M}_0^*)$ to the cross-section. Crucially, this correction is 
    purely kinematical in nature: it arises from retaining subleading terms in the kinematics 
    of the existing eikonal diagrams, and does not introduce any new Feynman structures beyond 
    those already present at leading order. This is precisely analogous to the treatment 
    of~\cite{Sadofyev:2021ohn}, in which sub-eikonal corrections of the same kinematical 
    character are retained without modifying the underlying diagrammatic content.
    
    To verify explicitly that the present kinematical corrections do not constitute genuine 
    sub-eikonal corrections in the sense of introducing new diagrammatic structures, we examine 
    the first diagram that does so — the crossed diagram. The crossed diagram first contributes 
    at $\mathcal{O}(1/P^{+4})$ in the interference term $\text{Re}(\mathcal{M}_2\mathcal{M}_0^*)$ 
    of the cross-section, which is two orders beyond the $\mathcal{O}(1/P^{+2})$ corrections to the cross-section
    retained here. Our kinematical corrections therefore remain well within the eikonal 
    approximation, and a fully consistent treatment of genuine sub-eikonal corrections,
    which would require the inclusion of the crossed diagram and related structures, lies 
    beyond the scope of the present work.

    % In this work, we also highlight the crucial role of unitarity. 
    % The preservation of unitarity, in light of the sub-Regge kinematical corrections is another 
    % Because we are calculating corrections to the momentum broadening distribution, our results must remain consistent with unitarity, as in the original GLV broadening calculation. 
    % Unitarity imposes constraints on sub-Regge kinematical corrections: simply 
    % Relaxing what may appear, at first glance, to be the eikonal approximation without care can lead to unitarity violation. 
    One may also wonder about the preservation of unitarity in light of the present sub-Regge corrections.
    We discuss this in \cref{Unitarity}, and show the computation of a sub-Regge kinematical result consistent with unitarity in \cref{SUBunitarity}.
    
    This paper is organised as follows. In \cref{setup}, we introduce the GLV framework for calculating the momentum-broadening distribution in \cref{setup}, including a detailed discussion of unitarity in \cref{Unitarity}. We derive the momentum broadening $\hat{q}$ in \cref{sec:qhat}. In \cref{approximations}, we review the approximations underlying the GLV formalism, discuss their physical motivation, and outline how these approximations are systematically relaxed. The main analytical results of the paper are presented in \cref{calculation}, with the full derivation provided in a \cref{app:AppendixA}. Finally, in \cref{Numerical}, we perform a numerical study and present several figures illustrating the impact of the corrections.

\section{Setup \label{setup}}

\subsection{Diagramatic Setup}

    We begin by reproducing the GLV setup 
    for ease of reference, before presenting modifications that allow for the calculation of the APL and sub-Regge kinematical corrections.

    Throughout, we denote two-dimensional transverse (to the motion of the parton) vectors in boldface as $ \pperpvec$, three-dimensional spatial vectors as $\pvec = (p_z,  \pperpvec)$, and four-vectors in Minkowski or light-cone coordinates as $p = (p^0,\pvec) = [p^0+p^z,\,p^0-p^z,\pperpvec]$ respectively. The magnitude of $\pperpvec$ and $\qperpvec$ are denoted as $p_{\perp} = |\pperpvec|$ and $q_{\perp} = |\qperpvec|$.

    A high-momentum parton is initially produced at $(t_0, z_0, \boldsymbol{x}_0)$ inside a finite QGP medium, where its interactions with target scattering centers are modeled via a Gyulassy–Wang Debye-screened potential\cite{Gyulassy:1999ig} with Fourier and color structure 
        \begin{equation}
            V_n     = V(q_n), e^{-i  \qperpvec_n \cdot \boldsymbol{x}_n}
                    =2\pi \delta(q^0)\, v(\vec{q}_n)\, e^{-i  \qperpvec_n \cdot \boldsymbol{x}_n}\, T_{a_n}(R) \otimes T_{a_n}(n),\\
        \end{equation}
    with the Yukawa-screened form
        \begin{equation}
            v(\vec{q}_n) = \frac{4 \pi \alpha_s}{\vec{q}_n^{\,2} + \mu^2}\,,
        \end{equation}
    and $q^0 = 0$, since the potential is static and does not vary with time.
    The color exchanges are handled using the appropriate SU($N_c$) generators: 
    $T_a(n)$ in the $d_n$-dimensional representation of the target, or $T_a(R)$ in the $d_R$-dimensional representation of the high-$p_T$ parent parton. 

     \begin{figure}[t]
                \centering
                \begin{subfigure}[]{0.18\textwidth}
                    \centering
                    \begin{tikzpicture}
                        \begin{feynman}
                            \vertex (a) [label={[yshift=+0.5cm]$P$}];
                            \vertex [right=1.5cm of a] (b);
                            \draw[pattern=north east lines] (a) circle (0.25);
                            \node [below=1.5cm of b] (b1);
                            \diagram* {
                                (a) -- [fermion, edge label=$p$] (b) ,                            
                            };
                        \end{feynman}
                    \end{tikzpicture}
                    \caption{$\mathcal{M}_0$}
                    \label{fig:no_scattering}
                \end{subfigure}
                \begin{subfigure}[]{0.28\textwidth}
                    \centering
                    \begin{tikzpicture}
                        \begin{feynman}
                            \vertex (a) [label={[yshift=+0.5cm]$P$}];
                            \vertex [right=1.3cm of a] (b);
                            \vertex [right=1.3cm of b] (c) [label={[yshift=+0.0cm]$p$}];
                            \node [crossed dot,below=1.5cm of b] (b1);
                            \draw[pattern=north east lines] (a) circle (0.25);
                            \diagram* {
                                (a) -- [fermion, edge label=$p-q$] (b) -- [fermion] (c),
                                (b1) -- [gluon, momentum'={[arrow shorten=0.2]$q$}] (b),
                            };
                        \end{feynman}
                    \end{tikzpicture}
                    \caption{$\mathcal{M}_1$}
                    \label{fig:single_scattering}
                \end{subfigure}
                \begin{subfigure}[]{0.48\textwidth}
                    \centering
                        \begin{tikzpicture}
                            \begin{feynman}
                                \vertex (a) [label={[yshift=+0.5cm]$P$}];
                                \vertex [right=2.0cm of a] (b);
                                \vertex [right=2.0cm of b] (c);
                                \vertex [right=2.0cm of c] (d) [label={[yshift=+0.0cm]$p$}];
                            
                                % Scattering centers
                                \node [crossed dot, below=1.5cm of b] (b1); % First scattering center
                                \node [crossed dot, below=1.5cm of c] (b2); % Second scattering center
                            
                                \draw[pattern={north east lines}] (a) circle (0.25);
                            
                                \diagram* {
                                  (a) -- [fermion, edge label=$p-q_1-q_2$] (b) -- [fermion, edge label'=$p-q_2$] (c) -- [fermion] (d),
                                  (b1) -- [gluon, momentum'={[arrow shorten=0.2]$q_1$}] (b),
                                  (b2) -- [gluon, momentum'={[arrow shorten=0.2]$q_2$}] (c)
                                };
                            \end{feynman}
                        \end{tikzpicture}
                    \caption{$\mathcal{M}_2$}
                    \label{fig:double_scattering}
                    \end{subfigure}
                    \caption{The relevant Feynman diagrams for momentum broadening at leading order in the opacity.}
                    \label{fig:diagrams}
            \end{figure}
    
    The relevant diagrams at leading order in the opacity are shown in \cref{fig:diagrams}. The initial transverse-momentum profile of jets produced by the hard scattering is highly collimated   : 
        \begin{align}
             \frac{dN^{(0)}}{d^3 \vec{p}} &= \frac{1}{2(2\pi)^3} |J(p)|^2 = f(E) \delta^{(2)} (\pperpvec). \label{eq:dN0}
        \end{align}
    Here $ f(E) $ encodes the momentum dependence of the parton, where $ E $ is the parton’s initial energy or equivalently in lightcone coordinates  $ \frac{\sqrt{2}}{2}P^+$. In the traditional GLV framework, one finds $ f(E) = 1/E^4 $. In \cref{FormalSubltleties}, this form is derived  in detail and several formal subtleties associated with this result are discussed.

        The full jet distribution, to first order in opacity, is given by (as in \cite{Gyulassy:2002yv})
            \begin{align}
                 \frac{dN^{(1)}}{d^3 \vec{p}} &= \int \rho (\Delta z)\int d^2 \qperp \Bar{\sigma}(\qperp) \left[ \frac{dN^{(0)}}{d^2 (\pperp - \qperp) dP^+} -  \frac{dN^{(0)}}{d^2 \pperp dP^+} \right],
            \end{align} 
            
    where, 
    
    \begin{align}
        \Bar{\sigma}(\qperp) \equiv \frac{d\sigma}{d^2 \qperp} = |v(\qperp)|^2.
    \end{align}
    
    The first order opacity in terms of matrix elements is taken to be,
    
            \begin{align}
                    \frac{dN^{(1)}}{d^3 \vec{p}} &=      \Bigg(
                    \frac{1}{d_T} \, \text{Tr}\langle \vert \mathcal{M}_1\vert ^2 \rangle
                    + \frac{2}{d_T}\, \text{Re}\,\text{Tr} \langle \mathcal{M}_2 \mathcal{M}_0^\ast \rangle 
                    \Bigg) , 
                \label{eq:distribution}
            \end{align}
    
    where $d_T$ is the dimension of the target color source. The single scattering matrix element is given by,
    
    \begin{align}
        \mathcal{M}_1 &= i e^{i p x_0} \int \frac{d^4 q}{(2 \pi)^4} J(p-q) \Delta(p-q) v(q) D(2p-q) \sum^N_{i=1} e^{-iq(x_i - x_0)} T^a(j) T^a(R) \\
        &= i e^{i p x_0} \int \frac{d^4 q}{(2 \pi)^4} J(p-q)\frac{1}{(p-q)^2 + i \epsilon}  2\pi \delta(q^0) \frac{4 \pi \alpha_s}{\vec{q}^2 + \mu^2} \sqrt{2}P^+  \sum^N_{i=1} e^{-iq(x_i - x_0)} \\& \times T^a(j) T^a(R) ,\label{eq:M1}
    \end{align}
    
     and the double scattering matrix element is given by,
    
    \begin{align}
        \mathcal{M}_2 &= i e^{i p x_0} \int \frac{d^4 q_1}{(2 \pi)^4} \int \frac{d^4 q_2}{(2 \pi)^4}J(p-q_1 - q_2) \frac{1}{(p-q_1)^2 + i \epsilon} \frac{1}{(p-q_1-q_2)^2 + i \epsilon} 2 P^{+2} \\& \times  2\pi \delta(q^0_1) \frac{4 \pi \alpha_s}{\vec{q_1}^2 + \mu^2} 2\pi \delta(q^0_2) \frac{4 \pi \alpha_s}{\vec{q_2}^2 + \mu^2} \sum^N_{i=1} e^{-iq(x_i - x_0)} \sum^N_{j=1} e^{-iq(x_j - x_0)} \\& \times T^a(j) T^b(i) T^a(j) T^b(R).\label{eq:M2}
    \end{align}

    The color traces and average over impact parameter give
            \begin{align}
                \text{Tr}\langle T_a(j) T_a(R)  T^{\ast}_a(j) T^{\ast}_a(R) \rangle 
                    &= C^2 (R)\, d_R 
                        = \frac{d_R}{d_A} C_2(R) C(R), \\
                        \langle e^{-i(\boldsymbol{q_{1}} + \boldsymbol{q_{2}})  \cdot \boldsymbol{b}} \rangle 
                    &= \frac{(2 \pi)^2}{A_{\perp}} \, \delta^{(2)} (\boldsymbol{q_{1}} + \boldsymbol{q_{2}}), \label{delta}
            \end{align}
    
    where $A_\perp$ is the transverse area of the colliding system.

         In the contact limit, the longitudinal separation between scatterings in the double scattering diagrams satisfies $z_2 - z_0 \rightarrow z_1 - z_0 \equiv \Delta z$.
         When performing the $q_2$ integration, the delta function enforces $\qperpvec_1 = - \qperpvec_2$ and thus implies ${q_{\perp,1}} = - q_{\perp,2}$.
         Consequently,
            \begin{equation}
                \sum_{j=1}^N e^{-i (q_{\perp,1} + q_{\perp,2})\Delta z} 
                = \sum_{j=1}^N e^0 
                = \sum_{j=1}^N 1 
                = N,
            \end{equation}            
        and the source function reduces as    
            \begin{equation}
                J(p -  \qperpvec_1 - \qperpvec_2) 
                = J(p - \qperpvec_1 + \qperpvec_1) 
                = J(p).
            \end{equation}
    
     Lastly, one is free to choose the distribution of scattering centres. An exploration of the effect of the choice of distribution is beyond the scope of the present work, but see \cite{Kolbe:2015rvk} for a discussion on different choices. We will employ the exponential distribution as in GLV:    
        \begin{align}
            \int \rho (\Delta z) &= \frac{2}{L} \int^\infty_0 (d\Delta z) e^{-\frac{2 \Delta z}{L}}.
        \end{align}

\subsection{Unitarity \label{Unitarity}}
        
    A crucial aspect of the present analysis is unitarity. 
    In the context of momentum broadening, preserving unitarity amounts to ensuring that the particle number remains conserved.
    Any correction to the broadening distribtution must, therefore, satisfy $\int d^3 \vec{p} dN^{(1)}/d^3 \vec{p} = 0$. Following \cite{Gyulassy:2002yv,Sadofyev:2021ohn}, 
    where $\sigma_0 = \int d^2 \qperp \, \Bar{\sigma}(\qperp),$ one may rewrite \cref{eq:distribution} as
    \begin{align}
        \frac{dN^{(1)}}{d^3 \vec{p}}
        &= \frac{N}{A_{\perp}} \frac{1}{d_A} C_2(R) C(R)
        \int \rho(\Delta z)
        \int \frac{d^2 \qperp}{(2 \pi)^2}
        \left( |J(p-\qperpvec)|^2 - |J(p)|^2 \right)
        \Bar{\sigma}(\qperp)
        \label{eq:Amplitude diff} \\
        &= \frac{N}{A_{\perp}} \frac{1}{d_A} C_2(R) C(R)
        \int \rho(\Delta z)
        \int \frac{d^2 \qperp}{(2 \pi)^2}
        |J(p-\qperpvec)|^2
        \left( \Bar{\sigma}(\qperp) - \sigma_0 \delta^{(2)}(\qperpvec) \right)
        \label{eq:Amplitide rewritten} \\
        &\approx \frac{N}{A_{\perp}} \frac{1}{d_A} C_2(R) C(R)
        |J(p)|^2
        \int \rho(\Delta z)
        \int \frac{d^2 \qperp}{(2 \pi)^2}
        \left( \Bar{\sigma}(\qperp) - \sigma_0 \delta^{(2)}(\qperpvec) \right).
    \end{align}
    
    The contributions from the single- and double-scattering diagrams are therefore incorporated through an effective shift in the elastic cross section, $\Bar{\sigma}(\qperp) \rightarrow \Bar{\sigma}(\qperp) - \sigma_0 \delta^{(2)}(\qperpvec)$. As a result, the jet broadening process is unitary, since $\int d^3 \vec{p} \, \frac{dN}{d^3 \vec{p}} = 0,$ which implies that the total number of final state particles is conserved. The effect of first-order scattering is solely to redistribute transverse momentum, leading to broadening rather than particle production or loss.
    In order for this unitary structure to be guaranteed, the broadening distribution must factorise as
    \begin{align}
        \frac{dN^{(1)}}{d^3 \vec{p}}
        = \frac{N}{A_{\perp}} \frac{1}{d_A} C_2(R) C(R)
        \int \rho(\Delta z)
        \int \frac{d^2 \qperp}{(2 \pi)^2}
        \left( |J(p-\qperpvec)|^2 - |J(p)|^2 \right)
        \mathcal{F}(\qperpvec),\label{eq:dN1factorized}
    \end{align}
    
    where $\mathcal{F}(\qperpvec)$ is the resultant function of the propagator and potential after the $\int dq_z$ has been performed. One may then perform a shift in integration variables, $ \pperpvec \rightarrow \qperpvec + \pperpvec $.
    Under this transformation, the single-scattering amplitude shifts as
    $J(p - \qperpvec) \rightarrow J(p)$,
    leading to an exact cancellation between the single- and double-scattering contributions.
    Consequently, $\int d^3 \vec{p}  \frac{dN^{(1)}}{d^3 \vec{p}} = 0, $
    ensuring the preservation of unitarity.
    To guarantee this factorised form, it is essential that the function $\mathcal{F}(\qperpvec)$ appearing in
    $\text{Tr}\langle |\mathcal{M}_1|^2 \rangle$
    be identical to that of $\mathcal{G}(\qperpvec)$ arising from
    $\text{Tr}\langle \mathcal{M}_2 \mathcal{M}_0^{\ast} \rangle$. The sum of the single- and double-scattering contributions must therefore take the form
    \begin{align}
        J(p - \qperpvec) \mathcal{F}(\qperpvec) - J(p) \mathcal{G}(\qperpvec)
    = \left( J(p - \qperpvec) - J(p) \right) \mathcal{F}(\qperpvec).
    \end{align}

    We have belaboured this well-known statement about unitarity because it is crucial in the computation of the next-to-leading order contributions to the Regge kinematics considered in the present work.
    On the one hand, unitarity provides conditions under which the expressions for the amplitudes involved are dramatically simplified.
    On the other hand, unitarity ensures that the cross-diagram does not contribute to the double-scattering amplitude, as described in \cref{crossed}

\subsection{Momentum Broadening ($\hat{q}$) Derivation \label{sec:qhat}} 

In this section, we outline the derivation for computing expectation values from the aforementioned momentum broadening distribution. As a primary application, we evaluate the mean squared transverse momentum, $\langle p^2_{\perp} \rangle$, from which the momentum broadening, $\hat{q}$ is subsequently computed.

    The expectation value is defined as,
    
    \begin{equation}
        \langle \cdots \rangle = \frac{\int d^2 p_{\perp} (\cdots) \frac{dN^{(1)}}{d^3 \vec{p}}}{\int d^2 p_{\perp} \frac{dN^{(0)}}{d^3 \vec{p}}},
    \end{equation}
    
    where $\frac{dN^{(0)}}{d^3 \vec{p}}$ is the initial distribution and $\frac{dN^{(1)}}{d^3 \vec{p}}$ is the distribution to first order in the opacity expansion. The initial transverse-momentum profile of jets produced by the hard scattering is highly collimated, hence the initial distribution is given by, \cref{eq:dN0}. Thus, the expectation value becomes, 
    
    \begin{align}
         \langle \cdots \rangle &= \frac{1}{f(E)} \int d^2 p_{\perp} (\cdots) \frac{dN^{(1)}}{d^3 \vec{p}}. \label{eq:expectation}
    \end{align}
    
    $\frac{dN^{(1)}}{d^3 \vec{p}}$ contains an integral over the internal momentum $\int dq$. After performing $\int dq_0 dq_z$ and computing the average over the impact parameter, one has $\frac{dN^{(1)}}{d^3 \vec{p}}$ is given by \cref{eq:dN1factorized}.
    Substituting \cref{eq:dN1factorized} for the first order in the opacity expansion into the expectation value \cref{eq:expectation}, we find,
    
    \begin{align}
        \langle \cdots \rangle &= \frac{1}{f(E)}  \frac{N}{A_{\perp}} \frac{1}{d_A} C_2(R) C(R)
        \int \rho(\Delta z) \int  d^2 p_{\perp} d^2 q_{\perp}  (\cdots) \\ & \times \left( |J(p-\qperpvec)|^2 - |J(p)|^2 \right)
        \mathcal{F}(\qperpvec).
    \end{align}
    
    Focusing on the source terms, we substitute the definitions of the source terms in terms of delta functions as highlighted in \cref{eq:dN0}. We find,
    
    \begin{align}
        \left( |J(p-\qperpvec)|^2 - |J(p)|^2 \right) &= f(E) \left( \delta^{(2)}(\pperpvec-\qperpvec) - \delta^{(2)}(\pperpvec) \right),
    \end{align}
    
    and so the expectation value becomes,
    
    \begin{align}
        \langle \cdots \rangle &=  \frac{N}{A_{\perp}} \frac{1}{d_A} C_2(R) C(R)
        \int \rho(\Delta z) \int  d^2 p_{\perp} d^2 q_{\perp}  (\cdots) \\& \times \left( \delta^{(2)}(\pperpvec-\qperpvec) - \delta^{(2)}(\pperpvec) \right)
        \mathcal{F}(\qperpvec).
    \end{align}

    Calculating the momentum broadening $\hat{q}$,
    
    \begin{align}
        \hat{q} &= \frac{\langle p^2_{\perp} \rangle}{L}\\
        &= \frac{1}{L} \frac{N}{A_{\perp}} \frac{1}{d_A} C_2(R) C(R)
        \int \rho(\Delta z) \int  d^2 p_{\perp} d^2 q_{\perp}  (p^2_{\perp}) \\ & \times \left( \delta^2(\pperpvec-\qperpvec) - \delta^2(\pperpvec) \right)
        \mathcal{F}(\qperpvec).
    \end{align}
    
    Performing the integral over $p^2_{\perp}$ and substituting $N/A_{\perp} = L/\lambda$, one has,
    
    \begin{align}
        \hat{q} 
        &= \frac{1}{L} \frac{N}{A_{\perp}} \frac{1}{d_A} C_2(R) C(R)
        \int \rho(\Delta z) \\ & \times \Bigg( \int  d^2 p_{\perp} d^2 q_{\perp}  (p^2_{\perp})  \delta^2(\pperpvec-\qperpvec) - \int d^2 p_{\perp} d^2 q_{\perp}  (p^2_{\perp})\delta^2(\pperpvec) \Bigg) 
        \mathcal{F}(\qperpvec)\\
        &= \frac{1}{\lambda}\frac{1}{d_A} C_2(R) C(R)
        \int \rho(\Delta z)\int  d^2 q_{\perp}  (q^2_{\perp}) \mathcal{F}.(\qperpvec)\label{eq:qhat}
    \end{align}

\section{Approximations \label{approximations}}

    The GLV formalism relies on a set of well-defined simplifying approximations that accurately describe parton energy loss in large collision systems such as nucleus–nucleus (AA) collisions. 
    However, the validity of these approximations becomes less certain in the context of small collision systems. 
    Consequently, examining the impact of systematically relaxing these approximations provides the central motivation for the present study.

    We will focus on two approximations in particular: the large separation distance and the broadening analogue of the large formation time approximations. The former was first relaxed in \cite{Kolbe:2015suq,Kolbe:2015rvk} where the impact of the second was realised. The impact of the second can also be seen in \cite{Sadofyev:2021ohn} and, as such, the present work presents a step towards reducing the severity of the first by introducing the second.\\

    \subsection{Large separation distance approximation }
    
    This approximation accounts for the spatial separation between the production point $z_0$ and the first scattering point $z_1$. 
    In principle, this approximation enforces the notion that the mean free path should be larger than the screening length of the scattering centre, an assumption that is crucial for the validity of the Gyulassy-Wang model:
        \begin{align}
        (z_{1} - z_{0}) \gg \lambda_{mfp} \implies \mu \Delta z \gg 1.
        \end{align} 
    
    In practice then, all terms proportional to $e^{-\mu \Delta z}$ are neglected under the large separation distance approximation.
    However, the statement that $ \mu \Delta z \gg 1$ enforces $\lambda_{mfp}\gg 1/\mu$ is not exactly correct, so that it must be possible to include contributions with small initial separation distances.
    The first attempt at computing an all-system-size correction to GLV energy-loss \cite{Kolbe:2015rvk,Kolbe:2015suq} relaxed precisely this approximation in the full radiative energy-loss calculation, keeping terms proportional to $e^{-\mu \Delta z}$. In the present work, we will also relax this approximation, calling the result the ``all-path-length'' (APL) result.\\

\subsection{Eikonal Approximation, Regge Kinematics and Sub-Regge 
Corrections}

In the high-energy limit, the initial energy of the parent parton is taken 
as the largest scale, leading to the eikonal hierarchy of scales:
\begin{align}
    k_{\perp} \ll P^+ \implies \frac{k_{\perp}^2}{P^+} \ll k_{\perp} \ll P^+.
\end{align}
In the radiative energy-loss case, this manifests as a ``large formation 
time'' approximation, in which the formation time of the radiated gluon
\begin{align}
    \frac{1}{\tau} = \frac{k_{\perp}^2}{x P^+} \ll \mu,
\end{align}
is taken to be much larger than the Debye screening length. In the present 
analysis, which focuses on momentum broadening (neglecting gluon radiation), 
an analogous broadening formation time naturally arises. Since 
$k_\perp \sim q_\perp$, the broadening analogue of the large formation time 
approximation is
\begin{align}
    \frac{1}{\tau_B} = \frac{q_\perp^2}{P^+} \ll \mu_{\perp}, 
    \label{eq:BFormationtime}
\end{align}
where $\mu_\perp^2 = \mu^2 + q_\perp^2$, recovering the standard eikonal 
separation of scales $q_\perp^2/P^+ \ll q_\perp \sim \mu_\perp \ll P^+$.

This hierarchy is made precise through the Regge condition $s \gg |t|$. 
For a parton with incoming momentum $P^\mu = [P^+, 0, \boldsymbol{0}]$ scattering 
off a Glauber gluon with momentum $q^\mu = [q_z/\sqrt{2},\, -q_z/\sqrt{2},\, 
\qperpvec]$, so that $q^0 = q^+ + q^- = 0$, with outgoing momentum 
$p^\mu = P^\mu - q^\mu$, one has
\begin{align}
    s = (P+q)^2, \qquad t = (P-p)^2 = q^2 = -q_z^2 - q_\perp^2,
\end{align}
so that $s \gg |t|$ becomes $(P+q)^2 \gg q_z^2 + q_\perp^2$. Explicitly,
\begin{align}
    s = 2P^+q^- = -\sqrt{2}\,P^+ q_z,
\end{align}
and since $q_z \sim q_\perp$, the Regge condition reduces to 
$P^+ \gg q_\perp$, precisely the eikonal condition. The standard GLV 
assignment
\begin{align}
    p^\mu = \left[P^+,\; \frac{q_z}{\sqrt{2}},\;\pperpvec\right],
\end{align}
is consistent with this limit, neglecting the longitudinal recoil of the 
scattered parton.

The improved kinematical assignment retains the leading correction in 
$q_\perp/P^+$,
\begin{align}
    p^\mu = \left[P^+ \Big(1 - \frac{q_z}{\sqrt{2}P^+} \Big),\; \frac{q_z}{\sqrt{2}},\; 
    \pperpvec\right],
\end{align}
which still satisfies $(P+q)^2 \gg q_z^2 + q_\perp^2$, remaining within 
the eikonal framework without requiring additional Wilson line structures 
or medium recoil. This correction is encoded in the $\gamma$-factor,
\begin{align}
    \gamma = \sqrt{1 - \frac{2q_\perp^2}{P^{+2}}},
\end{align}
which arises naturally during the calculation, with $\gamma \to 1$ as $q_\perp/P^+ \to 0$. In this way, $\gamma$ 
gives precise meaning to the broadening formation time $\tau_B$: the 
condition $\gamma \neq 1$ defines the kinematical regime in which 
sub-Regge corrections become important. The present work incorporates 
these effects by extending the kinematic range to
\begin{align}
    \mu \leq q_\perp \leq \frac{\sqrt{2}}{2} P^{+},
\end{align}
significantly enlarging the domain of applicability compared to the 
conventional Regge treatment. There are some formal subtleties associated 
with the sub-Regge correction at higher orders, including the role of the 
cross-diagram in the double scattering amplitude (addressed in 
\cref{crossed}) and the slow variation of the source term $J(p_\perp)$ 
with $p_\perp$ (demonstrated to hold up to order $1/(P^+)^4$ in 
\cref{FormalSubltleties}).

\section{Calculation \label{calculation}}

    To fully investigate the effects of relaxing the large separation distance and Regge approximations we revert back to where  approximations were first used; the kinematics. In the case of the APL correction we use the original Regge kinematics as the large seperation distance approximation is only applied when computing the matrix elements. In the case of the of the sub-Regge correction we keep a $q_z$ factor that is neglected in the standard Regge kinematics. The kinematics are summarized in \cref{tab:kin}.

    \begin{table}[t]
    \centering
    \renewcommand{\arraystretch}{1.6}
    \begin{tabular}{|p{0.18\textwidth}|p{0.78\textwidth}|}
    \hline
    \textbf{Approximation Scheme} & \textbf{Kinematics} \\
    \hline
    GLV & 
    \parbox[t]{\hsize}{
    $P = [P^+,0^-,\mathbf{0}] = \left(\frac{1}{\sqrt{2}}P^+, \frac{1}{\sqrt{2}}P^+, \mathbf{0}\right)$,\\
    $q = \left[\frac{1}{\sqrt{2}}q_z, -\frac{1}{\sqrt{2}}q_z, \pperpvec\right] = (0_0,q_z,\pperpvec)$,\\
    $p = \left[P^+, -\frac{1}{\sqrt{2}}q_z, \pperpvec\right] = (p_0, p_z, \pperpvec)$
    } \\
    \hline
    GLV+APL & Same as GLV \\
    \hline
    (GLV)$_{\text{SUB}}$ & 
    \parbox[t]{\hsize}{
    $P = [P^+,0^-,\mathbf{0}] = \left(\frac{1}{\sqrt{2}}P^+, \frac{1}{\sqrt{2}}P^+, \mathbf{0}\right)$,\\
    $q = \left[\frac{1}{\sqrt{2}}q_z, -\frac{1}{\sqrt{2}}q_z, \pperpvec\right] = (0,q_z,\pperpvec)$,\\
    $p = \left[P^+ + \frac{1}{\sqrt{2}}q_z, -\frac{1}{\sqrt{2}}q_z, \pperpvec\right] = (p_0, p_z, \pperpvec)$
    } \\
    \hline
    (GLV+APL)$_{\text{SUB}}$ & Same as (GLV)$_{\text{SUB}}$ \\
    \hline
    \end{tabular}
    \caption{Kinematics for different approximation schemes to momentum broadening considered in the present work.}
    \label{tab:kin}
    \end{table}

   When computing the matrix elements, the modified kinematics introduces corrections 
to the poles of the propagator, which now take the form
\begin{equation}
    q^{\pm}_z = -\frac{\sqrt{2}}{2}P^+\left(1 \pm \gamma\right) \pm i\epsilon,
\end{equation}
where we introduce the \textit{Sub-Regge factor} $\gamma \equiv \sqrt{1 - 2q_\perp^2/P^{+2}}$. 
In the standard GLV calculation, only the leading term in the Taylor expansion of 
$\gamma$ is retained. Although $\gamma$ encodes the full kinematical correction to 
all orders in $1/P^+$, the present calculation is self-consistent only up to 
$\mathcal{O}(1/P^+)$ at the level of the amplitude, as the correction is purely kinematical 
and introduces no new Feynman structures~\cite{Sadofyev:2021ohn}.

The fact that no additional Feynman diagrams contribute is confirmed by examining the two leading sources of higher-order contributions. 
The full source amplitude (see \cref{FormalSubltleties}) takes the form 
$J(p) \sim 1/(E + \qperp)^4$, generating corrections only at 
$\mathcal{O}(1/P^{+4})$. Similarly, the crossed diagram — identical to 
\cref{fig:double_scattering} but with crossed incoming gluons — reproduces 
the contact limit result, but contributes at $\mathcal{O}(1/P^{+4})$ in the 
well-separated case, three orders beyond what is retained here. A fully consistent 
treatment of these sub-eikonal contributions lies beyond the scope of the present work.

    Using the standard Regge kinematics for the GLV and $(\mathrm{GLV} + \mathrm{APL})$ frameworks, alongside the sub-Regge kinematics for $(\mathrm{GLV} )_{\mathrm{SUB}}$ and $(\mathrm{GLV} + \mathrm{APL})_{\mathrm{SUB}}$, we compute the momentum broadening distribution. The results are summarized in \cref{tab:dis}, details are given in \cref{app:AppendixA}. Notably the results from $ \text{Tr}\langle |\mathcal{M}_1|^2 \rangle $ are the same as the results from $ \text{Re}(\text{Tr}\langle \mathcal{M}^\ast_0 \mathcal{M}_2 \rangle)$, differing only by a crucial factor of $\frac{1}{2}$ and a shift in the amplitude.

  \begin{table}[t]
    \centering
    \renewcommand{\arraystretch}{1.6}
    \begin{tabular}{|p{0.18\textwidth}|p{0.78\textwidth}|}
    \hline
    \textbf{Approximation Scheme} & \textbf{$dN^{(1)}/d^3 \vec{p}$} \\
    \hline
    
    GLV & 
    $\begin{aligned}[t]
    \displaystyle
    \frac{N}{A_\perp} & \frac{1}{d_A}  C_2(R) C(R) (4\pi\alpha_s)^2  \int  \rho (\Delta z) \int \frac{d^2 \qperp}{(2\pi)^2} \\
    & \times \Big(|J(p-\qperpvec)|^2 - |J(p)|^2 \Big) \frac{1}{\muperp^4}
    \end{aligned}$ \\
    \hline
    
    GLV+APL & 
    $\begin{aligned}[t]
    \displaystyle
    \frac{N}{A_\perp} & \frac{1}{d_A}  C_2(R) C(R) (4\pi\alpha_s)^2  \int  \rho (\Delta z) \int \frac{d^2 \qperp}{(2\pi)^2} \\
    & \times \Big(|J(p-\qperpvec)|^2 - |J(p)|^2 \Big)
    \frac{1}{\muperp^4} \left( 1 - \tfrac{1}{2} e^{-\muperp \Delta z} \right)^2
    \end{aligned}$ \\
    \hline
    
    $(\mathrm{GLV})_{\mathrm{SUB}}$ & 
    $\begin{aligned}[t]
    \displaystyle
    \frac{N}{A_\perp} & \frac{1}{d_A}  C_2(R) C(R) (4\pi\alpha_s)^2  \int  \rho (\Delta z) \int \frac{d^2 \qperp}{(2\pi)^2} \\
    & \times \Big(|J(p-\pperpvec)|^2 - |J(p)|^2 \Big)
    \frac{4}{(1+\gamma)^2} \frac{1}{\muperp^4}
    \end{aligned}$ \\
    \hline
    
    $(\mathrm{GLV}+\mathrm{APL})_{\mathrm{SUB}}$ & 
    $\begin{aligned}[t]
    \displaystyle
    \frac{N}{A_\perp} & \frac{1}{d_A}  C_2(R) C(R) (4\pi\alpha_s)^2  \int  \rho (\Delta z) \int \frac{d^2 \qperp}{(2\pi)^2} \\
    & \times \Big(|J(p-\qperpvec)|^2 - |J(p)|^2 \Big) \frac{4}{(1+\gamma)^2} \frac{1}{\muperp^4} \left( 1 - \tfrac{1}{2} e^{-\muperp \Delta z} \right)^2
    \end{aligned}$ \\
    \hline
    
    \end{tabular}
    \caption{$dN^{(1)}/d^3 \vec{p}$ for different approximation schemes to momentum broadening considered in this work.}
    \label{tab:dis}
\end{table}

     In the case of the (GLV +APL) corrections, the correction introduces $(1 - \frac{1}{2}e^{-\muperp \Delta z})^2$ as is expected from \cite{Kolbe:2015rvk}. If one applies $e^{-\muperp \Delta z} \rightarrow 0 $, to APL, one recovers the original GLV result. Interestingly, the inclusion of the APL correction suggests there would be less broadening in small system sizes when $\muperp \Delta z$ is no longer $ \gg 1$. The exponential correction factor will dominate and suppress the broadening compared to original GLV.
    
    The sub-Regge kinematical correction to $dN^{(1)}/d^3 \vec{p}$ takes the form $4/(1 + \gamma)^2$. The introduction of the $\gamma$-factor allows for broadening effects when $\qperp \sim P^+$ and $\gamma <1$. The overall correction is proportional to $1/\gamma^2$. Since $\gamma \leq 1$,  the correction becomes $1/\gamma^2 \geq 1$. The inclusion of the sub-Regge kinematical correction suggests more broadening at large exchanged momentum compared to the initial incoming longitidual momentum of the parton. In the Regge limit when $\qperp \ll P^+ $, $\gamma \rightarrow 1$, and one recovers the GLV result.
    
    We also present a combined APL and sub-Regge kinematical correction to GLV. The $(\mathrm{GLV} + \mathrm{APL})_{\mathrm{SUB}}$ calculation is done using the sub-Regge  kinematics which leads to the $\gamma$-factor, in the form $4/(1 + \gamma)^2$, which is the same as the purely sub-Regge case. 
    The $(\mathrm{GLV} + \mathrm{APL})_{\mathrm{SUB}}$ calculation also includes the APL  terms $(1 - \frac{1}{2} e^{- \muperp \Delta z} )^2$. The combined result may be reduced to the individual corrections in the appropriate limits.

     After computing the momentum broadening distributions as outlined in \cref{tab:dis}, we proceed to calculate $\hat{q}$ using \cref{eq:qhat}. The results are summarized in \cref{tab:qhat}.

\begin{table}[t]
    \centering
    \renewcommand{\arraystretch}{1.6}
    \begin{tabular}{|p{0.18\textwidth}|p{0.78\textwidth}|}
    \hline
    \textbf{Correction Scheme} & \textbf{$\hat{q}$} \\
    \hline

    GLV &
    $\begin{aligned}[t]
    \displaystyle
    \frac{1}{\lambda} \frac{1}{d_A} & C_2(R) C(R) (4\pi\alpha_s)^2 \int \rho(\Delta z) \int \frac{d^2 \qperp}{(2\pi)^2} \\
    & \times \qperp^2 \frac{1}{\muperp^4}
    \end{aligned}$
    \\
    \hline

    GLV+APL &
    $\begin{aligned}[t]
    \displaystyle
    \frac{1}{\lambda} \frac{1}{d_A} & C_2(R) C(R) (4\pi\alpha_s)^2 \int \rho(\Delta z) \int \frac{d^2 \qperp}{(2\pi)^2} \\
    & \times \qperp^2 \frac{1}{\muperp^4} \left( 1 - \tfrac{1}{2} e^{-\muperp \Delta z} \right)^2
    \end{aligned}$
    \\
    \hline

    $(\mathrm{GLV})_{\mathrm{SUB}}$ &
    $\begin{aligned}[t]
    \displaystyle
    \frac{1}{\lambda} \frac{1}{d_A} & C_2(R) C(R) (4\pi\alpha_s)^2 \int \rho(\Delta z) \int \frac{d^2 \qperp}{(2\pi)^2} \\
    & \times \qperp^2 \frac{4}{(1+\gamma)^2} \frac{1}{\muperp^4}
    \end{aligned}$
    \\
    \hline

    $(\mathrm{GLV}+\mathrm{APL})_{\mathrm{SUB}}$ &
    $\begin{aligned}[t]
    \displaystyle
    \frac{1}{\lambda} \frac{1}{d_A} & C_2(R) C(R) (4\pi\alpha_s)^2 \int \rho(\Delta z) \int \frac{d^2 \qperp}{(2\pi)^2} \\
    & \times \qperp^2 \frac{4}{(1+\gamma)^2} \frac{1}{\muperp^4} \left( 1 - \tfrac{1}{2} e^{-\muperp \Delta z} \right)^2
    \end{aligned}$
    \\
    \hline

    \end{tabular}
    \caption{$\hat{q}$ for different approximation schemes to momentum broadening considered in this work.}
    \label{tab:qhat}
\end{table}

\section{Discussion and Numerical investigation \label{Numerical}}

    To investigate the full impact of the $(\mathrm{GLV} + \mathrm{APL})$, $(\mathrm{GLV})_{\mathrm{SUB}}$ and $(\mathrm{GLV} + \mathrm{APL})_{\mathrm{SUB}}$ corrections, we present a numerical analysis of the resulting momentum broadening distributions as a function of the broadened momentum $\pperp$. In all cases, we adopt the parameters $\alpha_s = 0.3$, $\mu = 0.5~\mathrm{GeV}$, $\lambda = 1~\mathrm{fm}$, an initial parton momentum $P^+ = 40~\mathrm{GeV}$, and a system size $L = 4~\mathrm{fm}$. The color factors used are $C_F = 4/3$ for the quadratic Casimir of the fundamental representation, $C_A = 3$ for the quadratic Casimir of the adjoint representation, and $d_A = 8$ for the dimension of the adjoint representation. We do not present results for an incoming gluon, as they differ from the quark case only by an overall color factor, which cancels in the ratio.

  \subsection{Momentum Broadening Distribution.}  

    To derive the transverse momentum distribution, we utilize a set of Fourier transforms to map the dynamics between momentum and coordinate space. The initial transform maps the scattering amplitudes from momentum-transfer space to impact-parameter space. A second Fourier transform then maps the unitarized coordinate-space distribution back to the momentum domain, yielding the final $p_\perp$ broadening spectrum. The resulting broadened distribution takes the form
\begin{align}
    \frac{dN^{(1)}}{d^3 \vec{p}} &= \frac{N}{A_{\perp}} \frac{1}{d_{A}}
       C_{2}(R) C(R) (4\pi \alpha_{s})^{2} \mathcal{F}(\pperpvec), \label{eq:Fourier}
\end{align}
where $\mathcal{F}(\pperpvec)$ encodes the analytic structure of the elastic cross section without resorting 
to the small impact parameter approximation. The latter approximation, which truncates 
the impact parameter space representation at quadratic order in $b$, yields the 
classic Moli\`ere Gaussian distribution in $p_\perp$~\cite{Gyulassy:2002yv}, and as we 
demonstrate below, can lead to substantial deviations from the GLV result.
    
     The transverse momentum transfer cannot fall below the Debye scale, giving the infared limit  $\pperp \geq \mu.$
    
    For the sub-Regge kinematical correction, the all-order factor $\gamma$ must be real. Its maximum value, $\gamma = 1$ corresponds to the maximum allowed momentum transfer $\qperp^{\mathrm{max}} = \sqrt{2}/2 \,P^{+}$ and provides an UV limit. Thus, the physically relevant transverse-momentum domain is
    \begin{align}
       \mu \leq \pperp \leq \frac{\sqrt{2}}{2} P^{+}. \label{eq:pperprange} 
    \end{align}
    
    \begin{figure}
            \centering
            \includegraphics[width=0.8\textwidth]{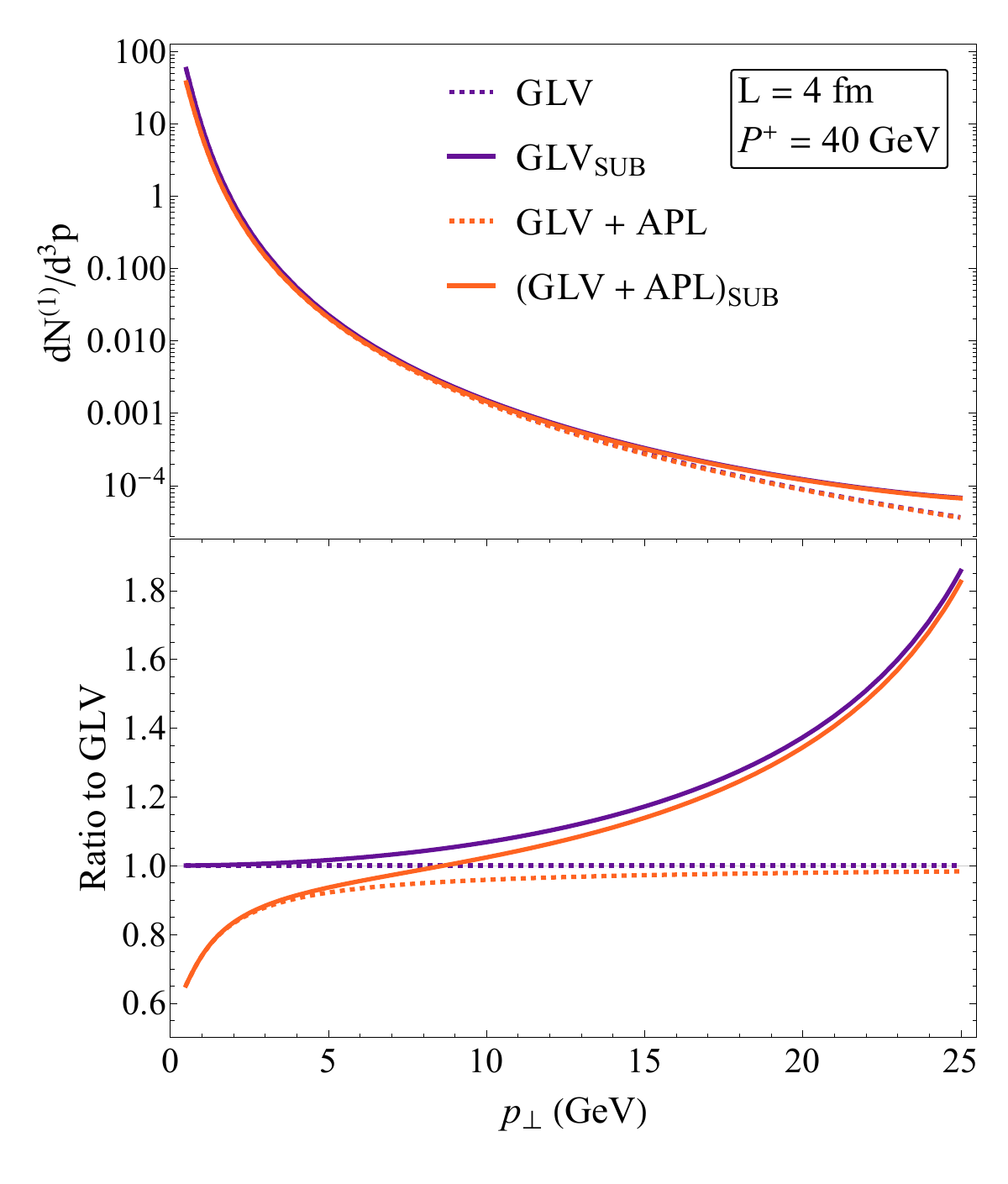}
        \caption{The broadening distribution $d^{(1)}N/d^3 \vec{p}$  for three correction schemes---(GLV + APL), sub-Regge kinematical $(\mathrm{GLV})_{\mathrm{sub}}$, and the combined correction $(\mathrm{GLV} + \mathrm{APL})_{\mathrm{sub}}$. Each corrected distribution is plotted both individually (top panel) and as a ratio to the unmodified GLV baseline (bottom panel), as a function of $\pperp$ for fixed initial parton momentum $P^+ = 40$ GeV and system size $L = 4$ fm.}
        \label{fig:Distribution}
    \end{figure}

        %\ik{use package subfigure package, see wikibooks \href{https://en.wikibooks.org/wiki/LaTeX/Floats,_Figures_and_Captions}{here}}
        
    % \begin{figure}[h]
    %   \centering
    %     \begin{minipage}[b]{0.48\textwidth}
    %         \centering
    %         \includegraphics[width=\linewidth]{Plots/L Plot v3.pdf}
    %         \caption{test caption\\label{fig:testLabel}}
    %     \end{minipage}\hfill
    %     \begin{minipage}[b]{0.48\textwidth}
    %         \centering
    %         \includegraphics[width=\linewidth]{Plots/Pp plot v3.pdf}
    %     \end{minipage}
    %     \caption{(a) The ratio of the GLV broadening distribution with all path-length corrections 
    % $(\mathrm{GLV} + \mathrm{APL})$ and the GLV broadening distribution with both 
    % all path-length and sub-Regge kinematical corrections $(\mathrm{GLV} + \mathrm{APL})_{\mathrm{sub}}$ 
    % to the original GLV distribution, plotted as a function of the transverse momentum $p_T$ 
    % for different system sizes $L$. 
    % (b) The ratio of the GLV broadening distribution with sub-Regge kinematical correction 
    % $(\mathrm{GLV})_{\mathrm{sub}}$ and the GLV broadening distribution with both 
    % all path-length and sub-Regge kinematical corrections $(\mathrm{GLV} + \mathrm{APL})_{\mathrm{sub}}$ 
    % to the original GLV distribution, plotted as a function of the transverse momentum $p_T$ 
    % for different initial momenta $P^+$.}
    %     \label{fig:L+Pp}
    % \end{figure}

\begin{figure}[t]
    \centering
    \begin{subfigure}[b]{0.49\textwidth}
        \includegraphics[width=\textwidth]{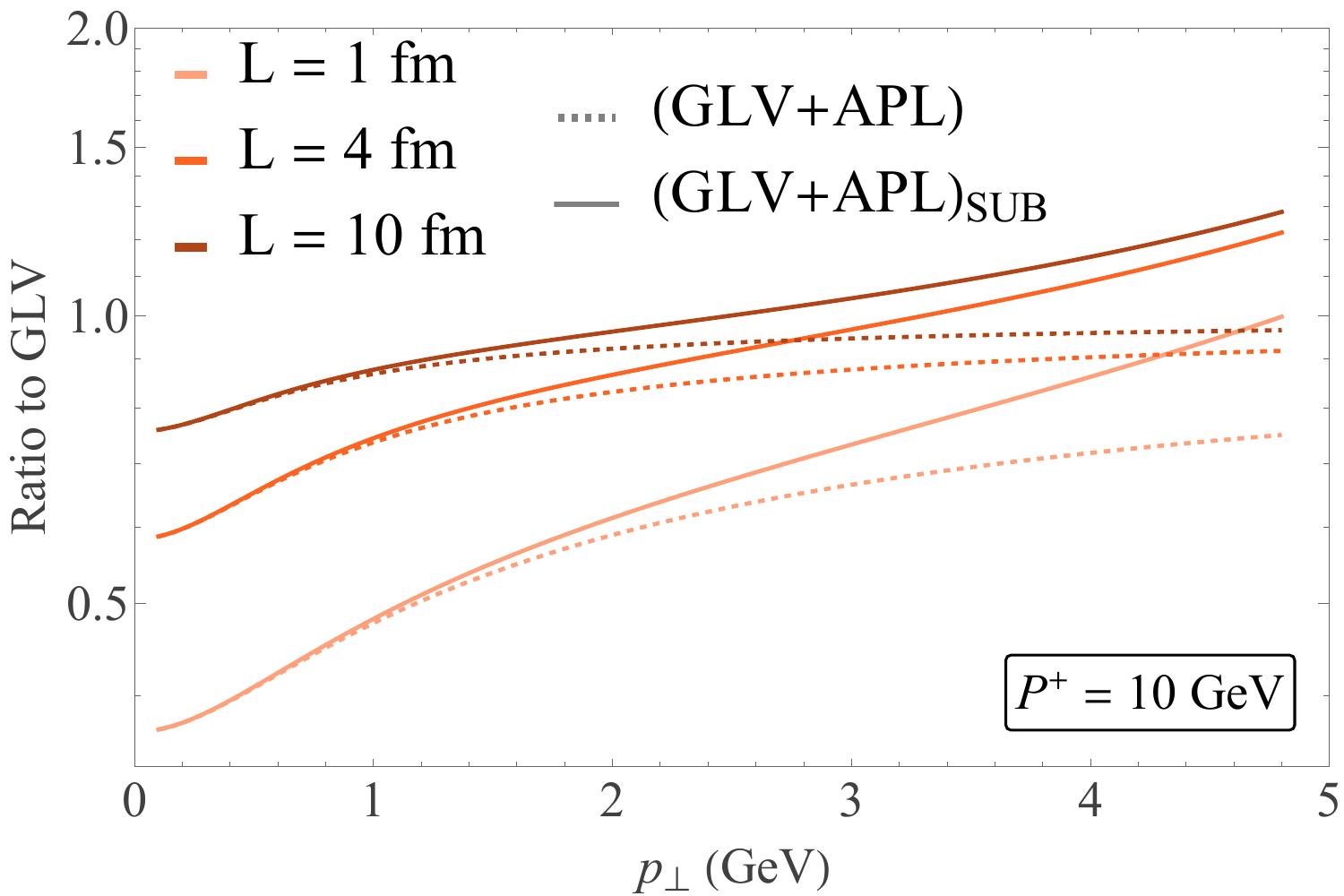}
        \caption{}
        \label{fig:LPlot}
    \end{subfigure}\hfill
    \begin{subfigure}[b]{0.49\textwidth}
        \includegraphics[width=\textwidth]{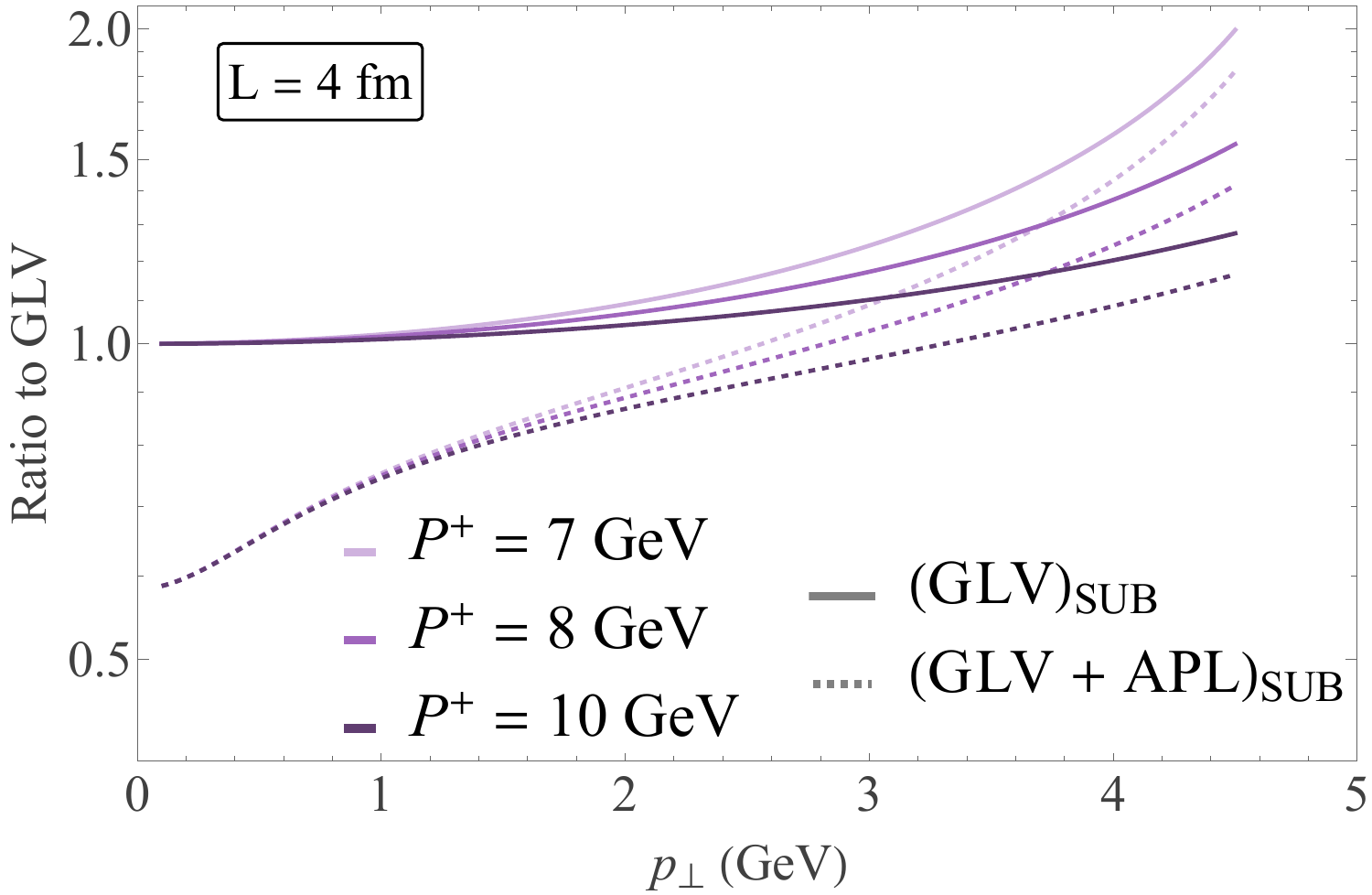}
        \caption{}
        \label{fig:PpPlot}
    \end{subfigure}
    \caption{Ratios of the modified broadening distributions---the all-path-length-corrected $(\mathrm{GLV} + \mathrm{APL})$, the sub-Regge kinematically corrected $(\mathrm{GLV})_{\mathrm{SUB}}$, and the combined $(\mathrm{GLV} + \mathrm{APL})_{\mathrm{SUB}}$ case---relative to the standard GLV baseline as a function of $\pperp$. Panel (a) illustrates the path length corrected results across system sizes $L = 1, 4, 10~\mathrm{fm}$ at a fixed initial parton momentum $P^+ = 10~\mathrm{GeV}$. Panel (b) illustrates the kinematical corrected results across initial momenta $P^{+} = 7, 8, 10~\mathrm{GeV}$ at a fixed system size $L = 4~\mathrm{fm}$.}
    \label{fig:LandPpplot}
\end{figure}
    
    The momentum broadening of the GLV model is plotted against 
    the transverse momentum $\pperp$ in \cref{fig:Distribution}, where the plot may directly be compared to the results presented in Fig.3 of \cite{Gyulassy:2002yv}. There is a suppression of the (GLV + APL) corrected distribution at low-$\pperp$. At high-$\pperp$ the effect is more obvious where there is an enhancement of the sub-Regge kinematical corrected curves. 
    
    To see the effects more clearly we plot the ratio of the the distribution with a correction to the standard GLV distribution in the bottom panel of \cref{fig:Distribution}. 
    In \cref{fig:Distribution}, at the low-$\pperp$ limit, there is a clear divergence of the (GLV + APL) corrected distribution relative to GLV, leading to a 
    steep fall-off near the lower bound $\muperp = \mu$. At low-$\pperp$, corresponding to small 
    $\muperp$, the correction term dominates and suppresses the GLV result. At high- 
    $\pperp$, when $\muperp$ is large, the APL correction tends to zero $(e^{-\muperp \Delta z} \rightarrow 0)$, and the 
    $(\mathrm{GLV} + \mathrm{APL})$ distribution converges to the original GLV result.
    
    The GLV distribution with sub-Regge kinematical corrections, $(\mathrm{GLV})_{\mathrm{SUB}}$, 
    diverges from the GLV result in the high $\pperp$ limit. In contrast, at low-$\pperp$, the sub-Regge kinematical 
    correction tends to zero, and $(\mathrm{GLV})_{\mathrm{SUB}}$ converges to GLV, 
    consistent with $\qperp/ P^+ \ll 1$ and $\gamma \rightarrow 1$.
    
    We also consider the combined APL and sub-Regge kinematical correction 
    $(\mathrm{GLV} + \mathrm{APL})_{\mathrm{SUB}}$. In the low-$\pperp$ limit, this 
    distribution approaches $(\mathrm{GLV} + \mathrm{APL})$, as the sub-Regge kinematical correction 
    becomes negligible. In the high-$\pperp$ limit, the distribution approaches 
    $(\mathrm{GLV})_{\mathrm{SUB}}$, as the APL correction vanishes. That is, suggesting 
    reduced broadening at low-$\pperp$ and enhanced broadening at high-$\pperp$ as compared to original GLV.
    
    To investigate the effect of APL correction, we present \cref{fig:LPlot}. 
    The two APL corrected distributions, $(\mathrm{GLV} + \mathrm{APL})$ 
    and $(\mathrm{GLV} + \mathrm{APL})_{\mathrm{SUB}}$ are shown, relative to GLV for three different 
    system sizes $L$ ($L = 1,4,10$ fm). After performing the integral over the distance of scattering 
    centres $\int d\Delta z$, the APL corrections scale as $\propto 1/(L \, \muperp)$. 
    At low-$\pperp$, which corresponds to small $\muperp$, the APL corrections dominate, smaller $L$ corresponds to a larger 
    correction factor, while larger $L$ reduces the correction and the result approaches GLV. At high-$\pperp$, which corresponds to large $\muperp$, the APL corrections converge to GLV, and 
    $(\mathrm{GLV} + \mathrm{APL})_{\mathrm{SUB}}$ converges to $(\mathrm{GLV})_{\mathrm{SUB}}$, 
    consistent with $1/(L \, \muperp) \to 0$.
    
    To investigate the effect of the sub-Regge kinematical correction, we present \cref{fig:PpPlot}. We plot the ratios of the two sub-Regge kinematical corrected distributions $(\mathrm{GLV})_{\mathrm{SUB}}$ and 
    $(\mathrm{GLV} + \mathrm{APL})_{\mathrm{SUB}}$ to GLV as a function of $\pperp$ for 
    three different initial parton momenta $P^+$ ($P^+ = 7,8,10$ GeV). As discussed, the 
    sub-Regge kinematical corrections dominate at large $\pperp$ as this is when the  $\gamma$-factor can no longer be approximated as 1.
    
    A smaller initial momentum $P^{+}$ leads to larger corrections within the same $\pperp$ range. The sub-Regge kinematical terms become relevant more rapidly for smaller $P^{+}$ because $\qperp$ approaches $P^{+}$ more quickly, causing $\gamma$ to be less than 1 sooner. At low-$\pperp$, the sub-Regge kinematical corrections vanish as $\qperp/P^{+} \ll 1$ and $\gamma \to 1$. Thus, 
    $(\mathrm{GLV})_{\mathrm{SUB}}$ converges to GLV and 
    $(\mathrm{GLV} + \mathrm{APL})_{\mathrm{SUB}}$ converges to $(\mathrm{GLV} + \mathrm{APL})$ 
    for all $P^+$, consistent with Regge approximation $\qperp / P^+ \ll 1$.
    
\begin{figure}[t]
        \centering
        % Left Subfigure
        \begin{subfigure}[b]{0.49\textwidth}
            \centering
            \includegraphics[width=\textwidth]{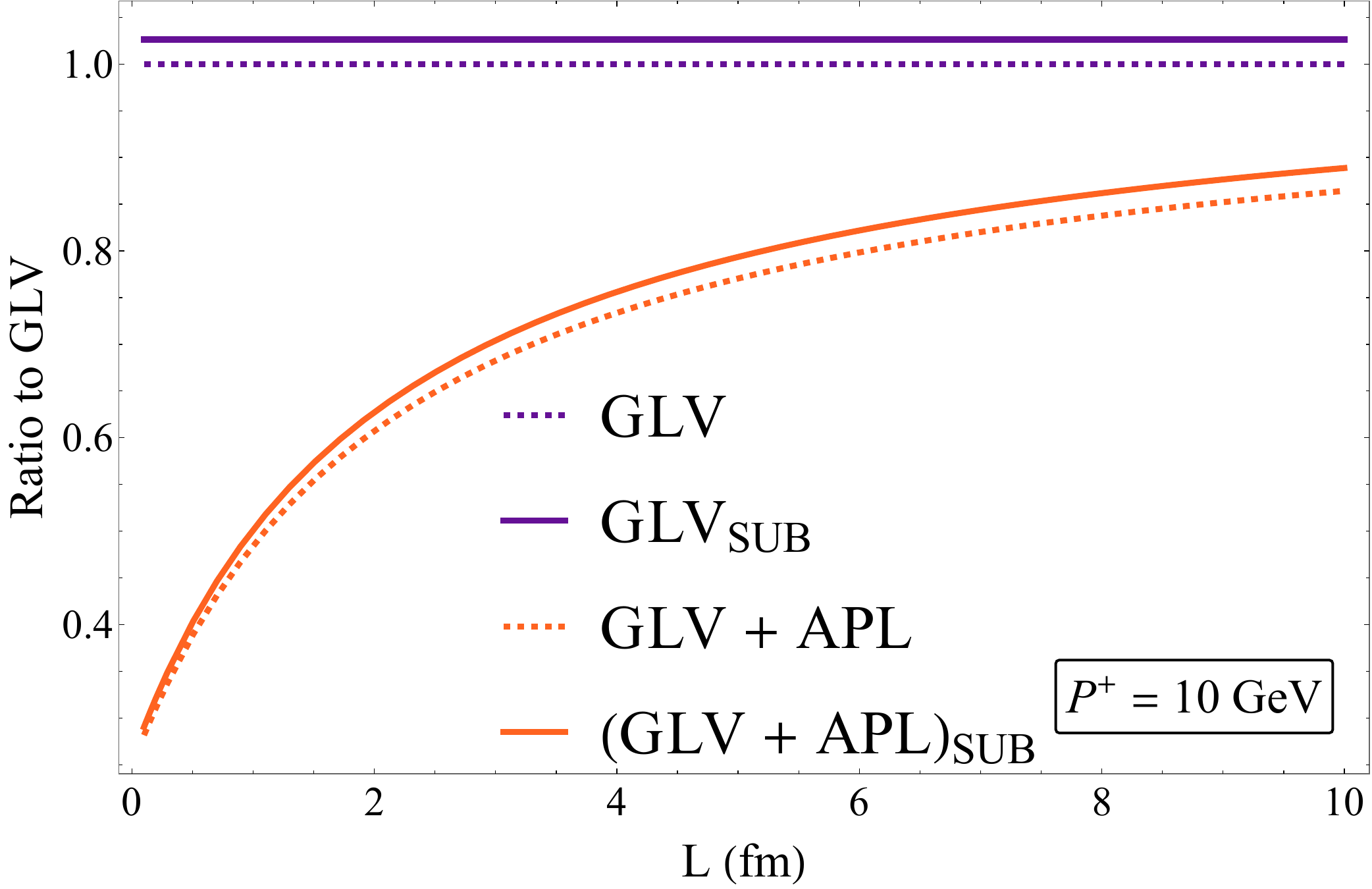}
            \caption{}
            \label{fig:vsL}
        \end{subfigure}
        \hfill % Adds horizontal stretching space between the plots
        % Right Subfigure
        \begin{subfigure}[b]{0.49\textwidth}
            \centering
            \includegraphics[width=\textwidth]{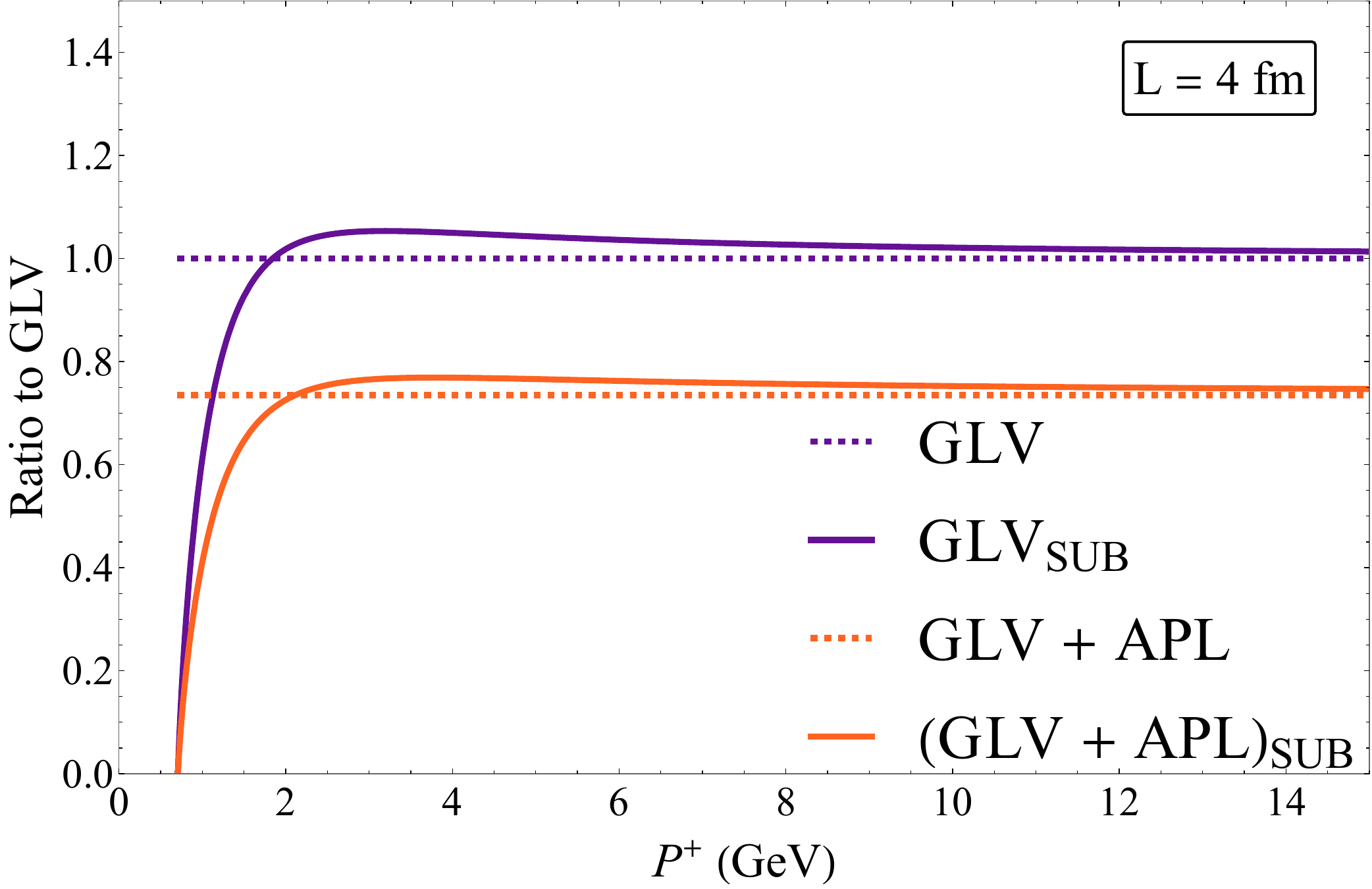}
            \caption{}
            \label{fig:vsPpplot}
        \end{subfigure}
        
        \caption{The ratios of the various corrected broadening distributions — the all path-length–corrected $(\mathrm{GLV} + \mathrm{APL})$, the sub-Regge kinematical corrected $(\mathrm{GLV})_{\mathrm{SUB}}$, and the combined all path-length and sub-Regge kinematical corrected $(\mathrm{GLV} + \mathrm{APL})_{\mathrm{SUB}}$ distributions — to the standard GLV result  as function of (a) system size L, for a fixed initial parton momentum of $P^+ = 10$ GeV, and (b) as a function of initial parton momentum $P^+$, for a fixed system size $L=4$ fm.  }
        \label{fig:combined_comparison}
    \end{figure}

    To investigate the relationship between the APL and sub-Regge kinematical corrections we present \cref{fig:vsL}. The ratios of (GLV + APL), $(\mathrm{GLV})_{\mathrm{SUB}}$, and
    $(\mathrm{GLV} + \mathrm{APL})_{\mathrm{SUB}}$ to the original GLV result as a function of $L$,
    with the exchanged momentum held fixed. The exchanged transverse momentum $\pperp$ has been numerically integrated over the range specified in \cref{eq:pperprange}. In the
    case of GLV and $(\mathrm{GLV})_{\mathrm{SUB}}$, which exhibit a linear $L$ dependence through
    $N/A_\perp = L/\lambda$, taking the ratio removes the $L$ dependence.
    Consequently, both GLV and $(\mathrm{GLV})_{\mathrm{SUB}}$ ratios to GLV are constant as
    functions of $L$.
    
    For the APL corrections, one observes that at large $L$, where the APL correction
    vanishes, the corresponding ratios converge to their respective baseline models.
    For small $L$, however, where the APL correction dominates, we observe a steep
    fall-off, indicating a suppression of the momentum broadening for small $L$
    compared to GLV and $(\mathrm{GLV})_{\mathrm{SUB}}$
    
    An interesting comparison is provided by (GLV + APL) and
    $(\mathrm{GLV} + \mathrm{APL})_{\mathrm{SUB}}$ . Upon introducing the sub-Regge kinematical correction to the
    APL contribution, we find that the sub-Regge kinematical correction mitigates the suppression induced by
    the APL term and leads to an enhancement of the momentum broadening. Since the
    sub-Regge kinematical correction reduces the APL contribution and becomes largely positive
    at $\pperp \sim P^+$, it may provide a possible resolution to the large
    negative corrections at high energies reported in \cite{Kolbe:2015rvk}.
    
    Lastly, we present \cref{fig:vsPpplot}, which shows the momentum broadening distributions for  $(\mathrm{GLV})_{\mathrm{SUB}}$, $(\mathrm{GLV}+\mathrm{APL})$, and $(\mathrm{GLV}+\mathrm{APL})_{\mathrm{SUB}}$ as functions of $P^+$, with the system size held fixed. The exchanged transverse momentum $\pperp$ has been numerically integrated over the range specified in \cref{eq:pperprange}.
    
    We observe that the $(\mathrm{GLV}+\mathrm{APL})$ result lies below the GLV curve, indicating a reduction in momentum broadening when the APL correction is included. In contrast, the inclusion of sub-Regge kinematical corrections enhances momentum broadening: $(\mathrm{GLV})_{\mathrm{SUB}}$ lies above the GLV result for all values of $P^+$. These trends demonstrate that while the APL correction suppresses momentum broadening, the sub-Regge kinematical correction has the opposite effect.
    
    A key result is that when both corrections are included, the sub-Regge kinematical contribution partially mitigates the suppression induced by the APL correction. This is reflected in the fact that $(\mathrm{GLV}+\mathrm{APL})_{\mathrm{SUB}}$ lies above $(\mathrm{GLV}+\mathrm{APL})$ for all $P^+$.This behavior is consistent with the trends observed in \cref{fig:Distribution} and lends additional support to the view that sub-Regge kinematical corrections may help alleviate the large negative contributions at high energies reported in \cite{Kolbe:2015rvk}.
    
    Finally, we note that the sub-Regge kinematical correction is particularly significant at lower values of $P^+$. In this regime, $\qperp \sim P^+$ and the $\gamma$-factor satisfies $\gamma \neq 1$, enhancing the importance of sub-Regge kinematical effects.

\subsection{Momentum Broadening ($\hat{q}$).}

\begin{figure}[t]\centering\includegraphics[width=0.8\linewidth]{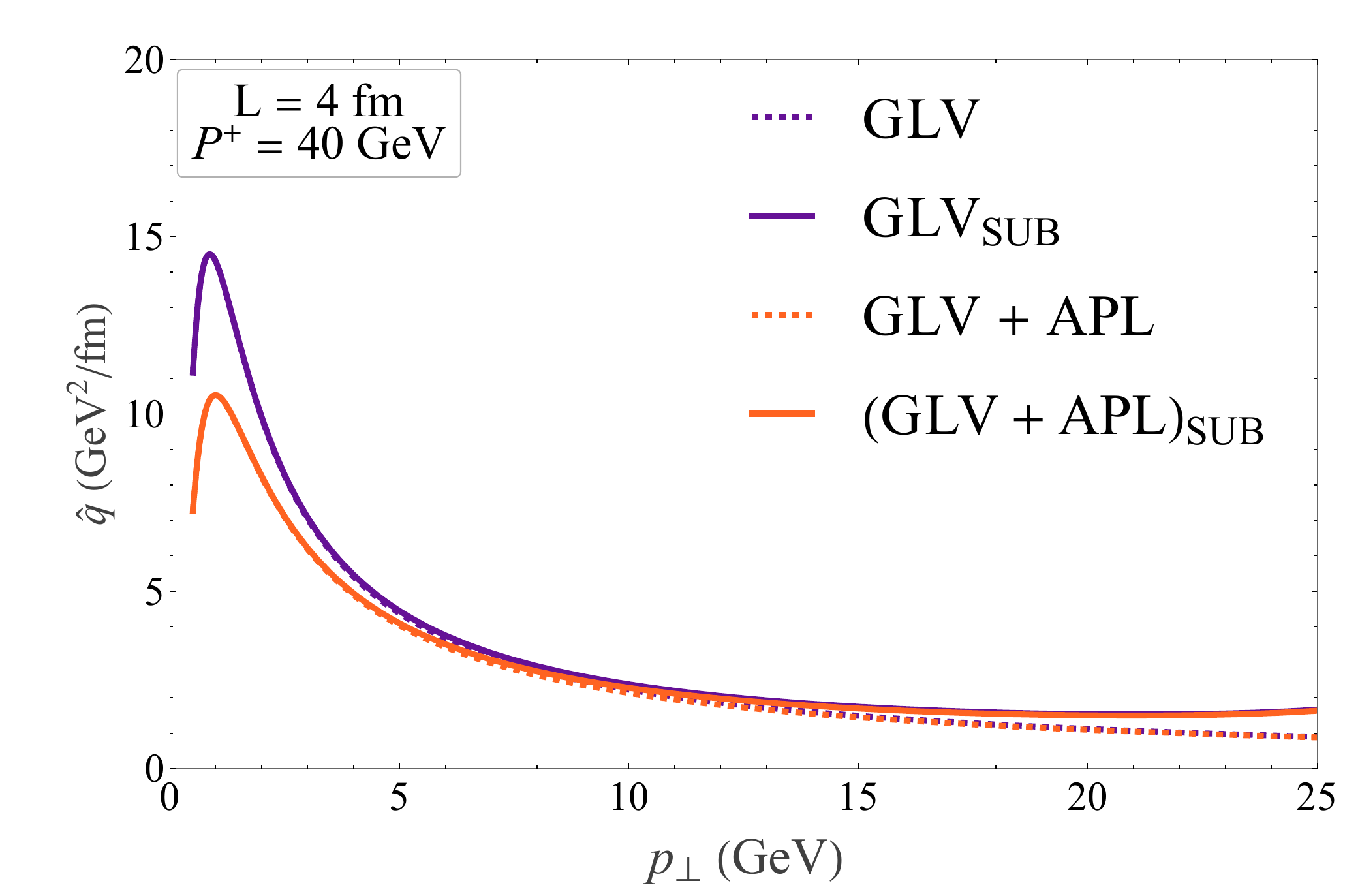}
\caption{The jet transport coefficient $\hat{q}$ evaluated as a function of $p_\perp$ for four distinct cases: the standard baseline $\hat{q}_{\mathrm{GLV}}$, the all-path-length-corrected $\hat{q}_{(\mathrm{GLV}+\mathrm{APL})}$, the sub-Regge kinematically corrected $\hat{q}_{(\mathrm{GLV})_{\mathrm{SUB}}}$, and the combined framework $\hat{q}_{(\mathrm{GLV}+\mathrm{APL})_{\mathrm{SUB}}}$. The kinematic parameters are fixed at an initial parton momentum $P^+ = 40~\mathrm{GeV}$ and a medium extent $L = 4~\mathrm{fm}$.}\label{fig:qhatvspperp}
\end{figure}

To investigate the quantitative impact of these corrections on the jet transport coefficient, we perform a numerical analysis. By applying the transformation procedure outlined in \cref{eq:Fourier}, $\hat{q}$ is evaluated as a function of the transverse momentum $\pperp$. The resulting distributions are presented in \cref{fig:qhatvspperp}.

Qualitatively mirroring the behavior of the underlying broadening distribution, the two non-APL-corrected formulations converge in the small-$\pperp$ regime, while the two APL-corrected variants converge in large-$\pperp$ regime. Notably, the APL-corrected curves lie systematically below the non-APL curves, illustrating the characteristic suppression induced by the all-path-length corrections at low transverse momentum. Furthermore, the sub-Regge kinematical corrections vanish in this infrared domain (low-$\pperp$), successfully recovering the expected asymptotic Regge limit. Conversely, in the ultraviolet (high-$\pperp$) limit, the sub-Regge kinematically corrected configurations asymptotically converge with each other, as do the non-sub-Regge baseline curves. The sub-Regge kinematically corrected curves remain systematically elevated relative to the non-corrected baselines, demonstrating that sub-Regge kinematics provide a distinct enhancement to momentum broadening at large $\pperp$.

\begin{figure}[t]
\centering
\begin{subfigure}[b]{0.48\textwidth}\centering\includegraphics[width=\textwidth]{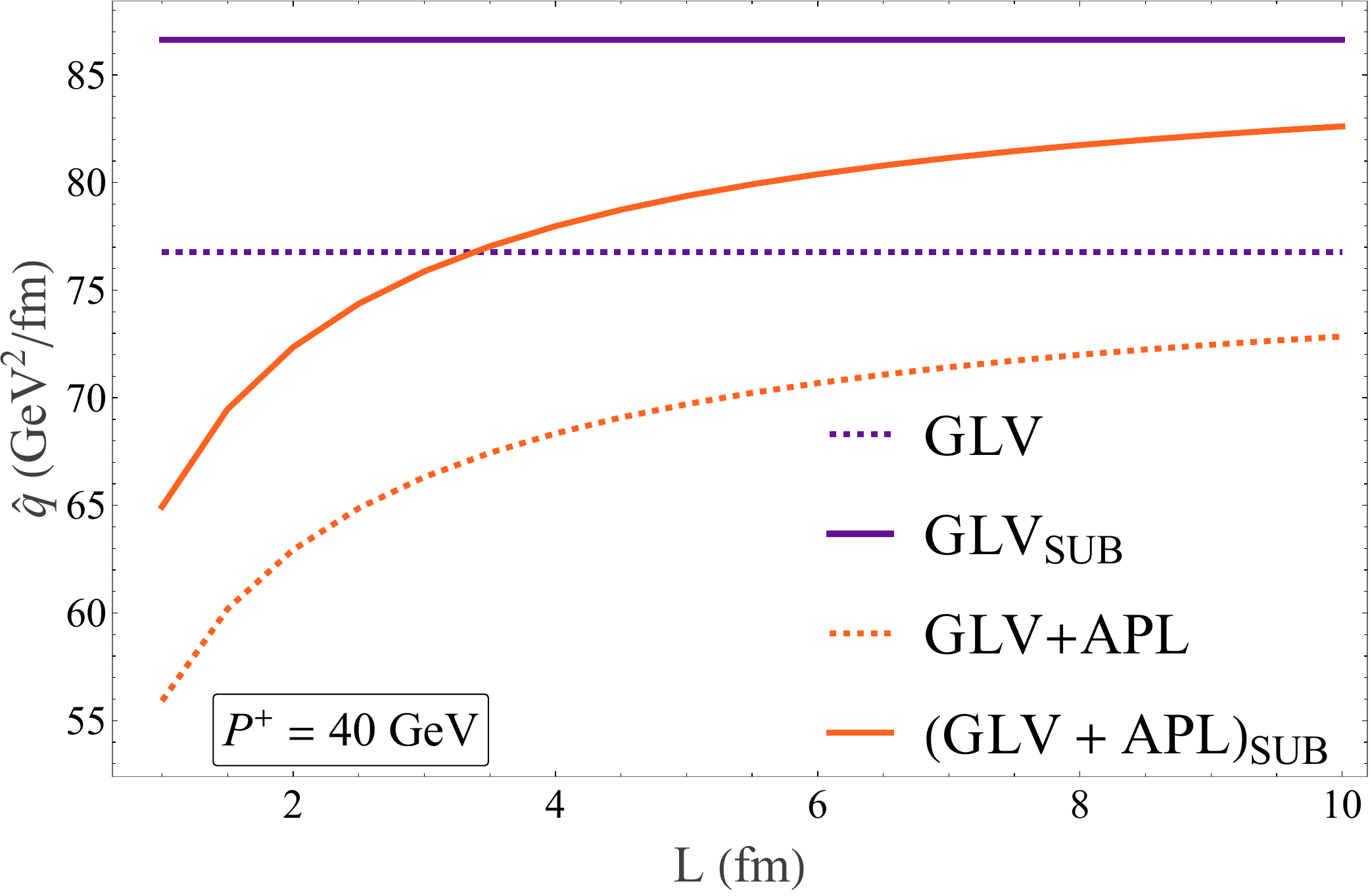}\caption{}\label{fig:qhatvsL}\end{subfigure}\hfill\begin{subfigure}[b]{0.49\textwidth}\centering\includegraphics[width=\textwidth]{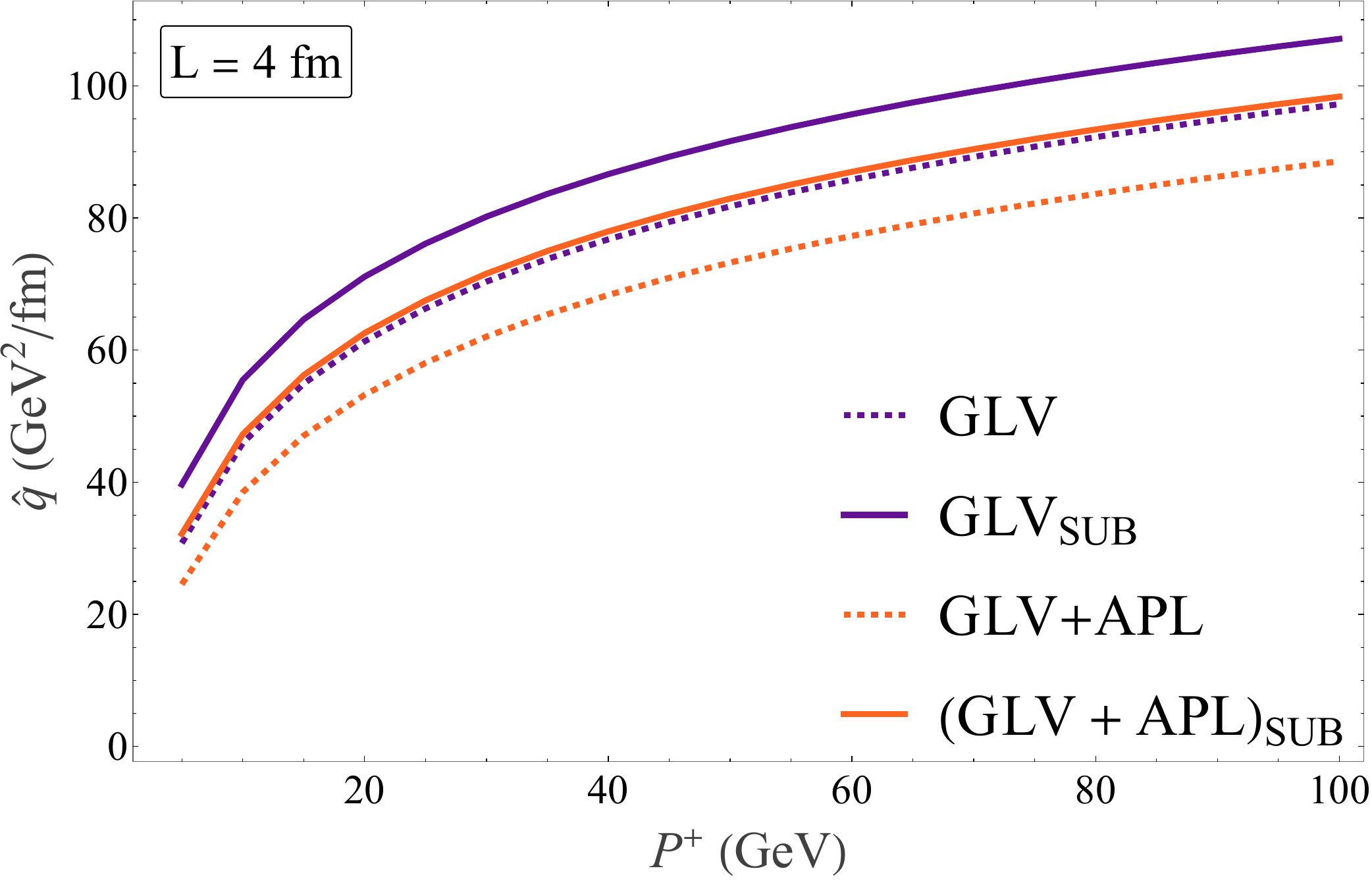 }
\caption{}
\label{fig:qhatvsPplus}
\end{subfigure}
\caption{The jet transport coefficient $\hat{q}$ plotted as a function of (a) the medium path length $L$, and (b) the initial longitudinal parton momentum $P^+$.}
\label{fig:qhat_combined}
\end{figure}

In \cref{fig:qhatvsL}, the integration over $\pperp$ was executed utilizing the kinematic bounds defined in \cref{eq:pperprange}. In the standard Regge and sub-Regge cases, the system size dependence $L$ from the opacity density ($L/\lambda$) is exactly canceled by the inverse scaling factor ($1/L$) inherent to the definition of the transport coefficient, $\hat{q} \equiv \langle p^2_{\perp} \rangle /L$. Consequently, both $\hat{q}_{\mathrm{GLV}}$ and $\hat{q}_{\mathrm{SUB}}$ remain strictly invariant with respect to the medium length $L$. Because the sub-Regge kinematic framework incorporates additional phase-space contributions to the broadening, it exhibits a rigid, uniform enhancement relative to the standard GLV baseline. In contrast, the APL-corrected coefficients introduce an explicit spatial dependence, scaling inversely with the medium length ($\propto 1/L$). In the thin-medium regime (small $L$), the APL corrections dominate the dynamics, resulting in a pronounced suppression of momentum broadening. For extended media (large $L$), the APL correction disapears leading to the asymptotic convergence of the $\hat{q}_{\mathrm{APL}}$ and $\hat{q}_{\mathrm{APLSUB}}$ curves toward their respective $\hat{q}_{\mathrm{GLV}}$ and $\hat{q}_{\mathrm{SUB}}$ limits.

Finally, we explore the dependence of the transport coefficient on the initial longitudinal jet momentum $P^+$, with the transverse phase-space integration restricted to the domain specified in \cref{eq:pperprange}. The results are shown in \cref{fig:qhatvsPplus}. While all four configurations exhibit qualitatively parallel trends across the momentum spectrum, $\hat{q}_{\mathrm{SUB}}$ remains elevated above $\hat{q}_{\mathrm{GLV}}$, reinforcing the conclusion that sub-Regge kinematics enhance momentum broadening universally across $P^+$. Crucially, $\hat{q}_{\mathrm{APLSUB}}$ is found to lie nearly degenerate with the standard baseline curve, $\hat{q}_{\mathrm{GLV}}$. This precise cancellation indicates that the enhancement from sub-Regge kinematical corrections completely counterbalances the suppression generated by the APL corrections. This interplay provides strong evidence for a viable field-theoretic resolution to the long-standing unphysical negative energy loss behaviors documented in previous literature~\cite{Kolbe:2015suq,Kolbe:2015rvk}.

\section{Conclusion}
    
We have evaluated the momentum broadening distribution to first order in the opacity expansion, $dN^{(1)}/d^3\vec{p}$, alongside the corresponding jet transport coefficient, $\hat{q}$, within the GLV formalism. This framework systematically incorporates all-path-length $(\mathrm{GLV} + \mathrm{APL})$, sub-Regge kinematical $(\mathrm{GLV})_{\mathrm{SUB}}$, and combined $(\mathrm{GLV} + \mathrm{APL})_{\mathrm{SUB}}$ corrections. Crucially, each modified formulation consistently reduces to the standard eikonal GLV baseline in the appropriate asymptotic limits.

The APL corrections induce a noticeable suppression of momentum broadening in the infrared (low-$p_\perp$) regime. This correction scales as $\propto 1/(L\,p_\perp)$, meaning it dominates within small or thin media and naturally vanishes as $L \to \infty$. Conversely, the sub-Regge kinematical corrections provide a distinct enhancement to momentum broadening at large $p_\perp$. This kinematic correction becomes pronounced when the transverse momentum transfer approaches the scale ($q_\perp \sim P^+$), and it vanishes in the high-energy Regge limit ($q_\perp/P^+ \ll 1$).

When both modifications are simultaneously integrated into the combined framework, the sub-Regge kinematical corrections partially mitigate the suppression generated by the APL terms across the entire parameter space of $p_\perp$, $L$, and $P^+$ investigated. Remarkably, in the integrated case of the jet transport coefficient, this mitigation is found to be essentially complete: $\hat{q}_{(\mathrm{APL+SUB})}$ deviates from the standard baseline GLV result by less than $5\%$ across the entire spectrum of initial parton momenta $P^+$ considered. This interplay suggests that the inclusion of sub-Regge kinematical corrections may offer a rigorous field-theoretic resolution to the problematic, unphysical negative energy-loss contributions at high energies documented in previous literature~\cite{Kolbe:2015suq, Kolbe:2015rvk}.

The momentum broadening results presented herein are formally valid up to order $\mathcal{O}(1/P^{+2})$. Extending this calculation to higher orders in the sub-eikonal expansion would necessitate accounting for non-trivial corrections to the initial source amplitude $J(p)$ as well as additional double-scattering diagrams, both of which enter at $\mathcal{O}(1/P^{+4})$. Consequently, the present framework establishes a systematic starting point for such higher-order sub-eikonal improvements, while simultaneously providing the requisite foundation for generalizing both the sub-Regge kinematical and all-path-length corrections to the full radiative energy-loss case.

\appendix
\renewcommand{\thesection}{\Alph{section}}
\crefalias{section}{appendix}

\section{Details of results calculated }\label{app:AppendixA}

In this section, we present a detailed derivation of the analytical results summarized in \cref{tab:dis}. In particular, we explicitly trace the GLV and $(\mathrm{GLV})_{\mathrm{SUB}}$ calculations from their respective kinematic foundations, as each framework relies upon a distinct kinematic scheme. We then briefly review the implementation of the all-path-length (APL) corrections, the exhaustive details of which can be found in Ref.~\cite{Kolbe:2015rvk}.To arrive at the expressions in \cref{tab:dis}, the phase-space integrations over the internal momentum transfer $q$ for the first-order amplitude $\mathcal{M}_1$, and over $q_1$ and $q_2$ for the second-order amplitude $\mathcal{M}_2$, must be evaluated as prescribed in \cref{eq:M1,eq:M2}. The presence of the static energy-conserving delta function, $\delta(q^0)$, renders the integration over $dq^0$ trivial, yielding unity. In what follows, we provide the explicit details regarding the contour evaluation of the longitudinal momentum integration, $\int dq_z$. Conversely, the remaining transverse momentum integrations, $\int d^2 \qperp$, are evaluated numerically according to the procedure outlined in \cref{Numerical}.Throughout this derivation, we adopt the standard $\sqrt{2}$ convention for light-cone coordinates. It is worth noting that this choice stands in contrast to the original GLV formulations, which traditionally employ the $2$ convention . In this representation, a generic four-vector $x^\mu = (x^0, x^z, \boldsymbol{x})$ is projected onto the longitudinal light-cone components defined by\begin{equation}x^+ = \frac{1}{\sqrt{2}}(x^0 + x^z) , \quad x^- = \frac{1}{\sqrt{2}}(x^0 - x^z) ,\end{equation}with the corresponding inverse transformations given by\begin{equation}x^0 = \frac{1}{\sqrt{2}}(x^+ + x^-) , \quad x^z = \frac{1}{\sqrt{2}}(x^+ - x^-) .\end{equation}

\subsection{Standard GLV calculation}

\subsubsection{Single Scattering}

To compute the momentum broadening distribution we first compute $\mathcal{M}_1$ from \cref{eq:M1}. The $q_z$ integral can be calculated analytically using the residue theorem. 
The integral to be computed is
\begin{align}
 I_1 &= \int \frac{dq_z}{2 \pi} \frac{1}{(p-q)^2 + i \epsilon} \frac{4 \pi \alpha_s}{q_z^2 + \muperp^2} e^{-iq_z \Delta z}.\label{eq:I1}
\end{align}

Using $ (p-q)^2 + i \epsilon = -q^2_z + \sqrt{2}P^+q_z + \qperp^2+i\epsilon = - (q_z - q^{(+)}_z)(q_z - q^{(-)}_z) +i\epsilon $ and $q_z^2 + \muperp^2 = (q_z + i \muperp)(q_z - i\muperp)$, we close the contour in the negative complex plane. The contributing poles are given by
\begin{align}
q^{(1)}_z &=  -\frac{\sqrt{2}}{2} \frac{\qperp^2}{P^{+}} - i \epsilon,\\
q^{(2)}_z &= -i\muperp . 
\end{align}

The residues associated with each pole are
\begin{align}
    \text{Res}[I_1,q^{(1)}_z] &= \lim_{q_z\rightarrow q^{(1)}_z} \frac{-1}{(p-q)^2+i\epsilon} \frac{4 \pi \alpha_s}{q_z^2 + \muperp^2} e^{-iq_z \Delta z} (q_z-q^{(1)}_z)\\
    &= \frac{-1}{(\frac{\sqrt{2}}{2} \frac{\qperp^2}{P^{+}} - \sqrt{2} P^+)} \frac{4 \pi \alpha_s}{(\frac{\sqrt{2}}{2} \frac{\qperp^2}{P^{+}})^2 + \muperp^2} e^{-i(-\frac{\sqrt{2}}{2} \frac{\qperp^2}{P^{+}}) \Delta z}.
\end{align}

Since $\qperp \ll P^+ $ and $ \muperp^2 = \qperp^2 + \mu^2 \implies \frac{\qperp^2}{P^+} \ll \qperp \approx \muperp $. Using these approximations, one has, in standard GLV, the term $ \frac{-\sqrt{2}}{2}\frac{\qperp^2}{P^+} -\sqrt{2} P^+ \approx -\sqrt{2} P^+$ and $ (\frac{\sqrt{2}}{2} \frac{\qperp^2}{P^{+}})^2 + \muperp^2 = (\frac{\sqrt{2}}{2} \frac{\qperp^2}{P^{+}} + i\muperp)(\frac{\sqrt{2}}{2} \frac{\qperp^2}{P^{+}} - i\muperp) \approx (i\muperp)(- i\muperp) = \muperp^2 $. Since $\frac{\sqrt{2}}{2} \frac{\qperp^2}{P^{+}} \ll 1 $ and $ e^{ix} \approx 1$ for $ x\ll1 $ therefore $e^{i\frac{\sqrt{2}}{2} \frac{\qperp^2}{P^{+}} \Delta z} \approx 1$ and so,
\begin{align}
    \text{Res}[I_1,q^{(1)}_z] &=  \frac{1}{\sqrt{2} P^+} \frac{4 \pi \alpha_s}{ \muperp^2}.
\end{align}

The pole from the propagator gives,
\begin{align}
    \text{Res}[I_1,q^{(2)}_z] &= \lim_{q_z\rightarrow q^{(2)}_z}\frac{1}{(p-q)^2+i\epsilon} \frac{4 \pi \alpha_s}{q_z^2 + \muperp^2} e^{-iq_z \Delta z} (q_z - q^{(2)}_z) \\
    &= \frac{-1}{{(-  i\muperp +\frac{\sqrt{2}}{2} \frac{\qperp^2}{P^{+}})}(-  i\muperp - \sqrt{2}P^+)} \frac{4 \pi \alpha_s}{( - 2 i\muperp) } e^{-\muperp \Delta z} . 
\end{align}

Simplifying using the arguments stated above, one has $ \implies -  i\muperp - \sqrt{2}P^+ \approx - \sqrt{2}P^+ $, and $\frac{\qperp^2}{P^+} \ll \qperp \approx \muperp \implies -  i\muperp +\frac{\sqrt{2}}{2} \frac{\qperp^2}{P^{+}} \approx -  i\muperp$. Therefore the residue becomes,

\begin{align}
    \text{Res}[I_1,q^{(2)}_z] &=\frac{-1}{{(-  i\muperp )}( - \sqrt{2}P^+)} \frac{4 \pi \alpha_s}{( - 2 i\muperp) } e^{-\muperp \Delta z} =  \frac{-4 \pi \alpha_s}{2 \sqrt{2} P^+ \muperp^2} e^{-\muperp \Delta z}. \label{I1:nonzero}
\end{align}

In the large separation distance approximation $\muperp \Delta z \gg 1$. Therefore $e^{-\muperp \Delta z} \approx0$ and one has,

\begin{align}
    \text{Res}[I_1,q^{(2)}_z] &= 0 . \label{I1:zero}
\end{align}

Using the sum of the residues, the integral is given by,
\begin{align}
    I_1 
    &= \frac{-i}{\sqrt{2} P^+} \frac{4 \pi \alpha_s}{ \muperp^2}.
\end{align}

Taking the complex conjugate, the single scattering contribution to the broadening distribution is therefore given by,
\begin{align}
    \text{Tr}\langle |\mathcal{M}_1|^2 \rangle &= \frac{N}{A_\perp} \frac{1}{d_A} C_2(R) C(R) (4\pi\alpha_s)^2  \int \rho (\Delta z) \int \frac{d^2 \qperp}{(2\pi)^2}
|J(p-\qperp)|^2 
\frac{1}{\muperp^4}.
\end{align}

\subsubsection{Double Scattering}

The first integral to be computed is,
\begin{align}
    I_2 &= \int \frac{dq_{1,z}}{2 \pi} v(q_1) \frac{1}{(p - q_1 -q_2)^2 + i \epsilon} e^{-iq_{1,z} \Delta z}.
\end{align}

 It is simplest to make the substitution $ q_3 = q_1 + q_2$ and $ dq_3 = dq_1$. Therefore the integral to be computed becomes,
\begin{align}
    I_2 &= \int \frac{dq_{3,z}}{2 \pi} \frac{4 \pi \alpha_s}{\vec{q_{1}}^2 + \mu^2} \frac{1}{(p - q_3)^2 + i \epsilon} e^{-iq_{3,z} \Delta z}. \label{eq:I2}
\end{align}

 $I_2$ takes the same form as the previous integral $I_1$. Therefore, if we close the the contour in the lower half of the complex plane the poles from the propagator and potential that contribute are given by,
\begin{align}
    q^{(1)}_{1,z} &= -\frac{\sqrt{2}}{2}\frac{q^2_{\perp,3}}{P^{+}} - q_{2,z} - i \epsilon,\\
    q^{(2)}_{1,z} &= -i\mu_{1},   
\end{align}

where $ \mu^2_1= q^2_{\perp,1} + \mu^2$. Applying the same approximations as before

\begin{align}
  {\text{Res}[I_2,q^{(1)}_{z,1}] = \frac{4 \pi \alpha_s}{ \sqrt{2} P^+}\frac{1}{q^2_{2,z}+ \mu^2_1} } ,
\end{align}

\begin{align}
     \text{Res}[I_2,q^{(2)}_{z,1}] 
     &= \frac{-4\pi \alpha_s}{2 \sqrt{2} i \mu_1 P^+ (q_{2,z}-i\mu_1)}e^{-i(q_{2,z}-i\mu_1)\Delta z}. \label{I2zero}
\end{align}

In the large separation distance approximation $\muperp \Delta z \gg 1$. Therefore $e^{-\muperp \Delta z} \approx0$ and one has 
\begin{align}
    \text{Res}[I_2,q^{(2)}_{z,2}] = 0. \label{I2:zero} 
\end{align}

The integral in \cref{eq:I2} is then given by
\begin{align}
    I_2 &= \frac{-i}{\sqrt{2} P^+} \frac{4 \pi \alpha_s}{ q^2_{2,z} + \mu^2_1}. \label{eq:I2result}
\end{align}

The next integral to be compted is,

\begin{align}
    I_3 &= \int \frac{dq_{2,z}}{2 \pi}  \frac{4 \pi \alpha_s}{ q^2_{2,z} + \mu^2_1}  \frac{4 \pi \alpha_s}{\vec{q}^2_2+\mu^2} \frac{1}{(p-q_2)^2 +i\epsilon} e^{-iq_{2,z}\Delta z'}. \label{eq:I3}
\end{align}

Where $\Delta z' \equiv z_2 - z_1$. Closing the contour in the negative half of the complex plane, the resulting three poles are:

\begin{align}
    q^{(1)}_{2,z} &=  -\frac{\sqrt{2}}{2} \frac{q^2_{2,z}}{P^{+}} -i \epsilon \label{pole1}, \\
     q^{(2)}_{2,z} &= -i \mu_2 \label{pole2}, \\
     q^{(3)}_{2,z} &= -i \mu_1 \label{pole3}.  
\end{align}

Computing the residues with the same approximations,

\begin{align}
    \text{Res}[I_3 I_2,q^{(1)}_{2,z}] = \frac{-i(4 \pi \alpha_s)^2}{ 2  \mu^2_1 \mu^2_2 P^{+2}}.
\end{align}

\begin{align}
    \text{Res}[I_3 I_2, q^{(2)}_{2,z}] = \frac{i(4 \pi \alpha_s)^2}{4 P^{+2} \mu^2_1(\mu^2_2-\mu^2_1)} e^{-\mu_1 \Delta z'}
\end{align}

In the contact limit $z_2 \rightarrow z_1 $, so $\Delta z' \rightarrow 0$

\begin{align}
    \text{Res}[I_3, q^{(3)}_{2,z}] &= \frac{i(4 \pi \alpha_s)^2}{4 P^{+2} \mu^2_2(\mu^2_1-\mu^2_2)} .
\end{align}

Combining \cref{eq:I2result} and \cref{eq:I3} gives

\begin{align}
    I_3 I_2 &= \frac{-2 \pi i}{2 \pi} \sum^2_{j=1} \text{Res}[I_3 I_2,q^{(j)}_z] \\
    &= \frac{-(4 \pi \alpha_s)^2}{2 P^{+2} } \Bigg( \frac{1}{\mu^2_1 \mu^2_2} - \frac{1}{2 \mu^2_1 ( \mu^2_2 - \mu^2_1)} - \frac{1}{2 \mu^2_2 ( \mu^2_1 - \mu^2_2)} \Bigg) \\
    &= \frac{-(4 \pi \alpha_s)^2}{2 P^{+2} } \Bigg( \frac{\cancel{(\mu^2_2 - \mu^2_1)}}{2 \mu^2_1 \mu^2_2 \cancel{(\mu^2_2 - \mu^2_1)}} \Bigg) = \frac{-(4 \pi \alpha_s)^2}{4 P^{+2} \mu^2_1 \mu^2_2}.
\end{align}

Applying the average over the impact parameter as per our discussion in \cref{delta}, which sets $ \mu_1 = \mu_2$  as well as $J(p - \qperpvec_1 - \qperpvec_2) = J(p)$, $I_3 I_2$ becomes, 

\begin{align}
    I_3 I_2 &= \frac{-(4 \pi \alpha_s)^2}{4 P^{+2} \muperp^4}.
\end{align}

The non-interacting matrix, \cref{fig:no_scattering}, is given by
\begin{align}
    \mathcal{M}^{\ast}_0 = -ie^{-ipx_0} J^{\ast}(p) .
\end{align}

The double scattering contribution to the broadening distribution is therefore given by
\begin{align}
    \text{Tr}\langle \mathcal{M}_2 \mathcal{M}^{\ast}_0 \rangle &= \frac{N}{A_\perp} \frac{1}{d_A} C_2(R) C(R) (4\pi\alpha_s)^2  \int \rho (\Delta z) \int \frac{d^2 \qperp}{(2\pi)^2}
|J(p)|^2 
\frac{-1}{2\muperp^4}.
\end{align}

Crucially we note the result of the double scattering contribution differs from the result of the single scattering by a factor of $- \frac{1}{2}$. Since the double scattering term has an overall factor of 2 when calculating the broadening distribution cancels the factor of $\frac{1}{2}$, resulting in the single and double scattering contributions only differs in a shift in the amplitude.

\subsubsection{The momentum broadening distribution}

Our result for the distribution \cref{eq:distribution} is then

\begin{align}
     \frac{dN^{(1)}}{d^3 \vec{p}} &=   \Bigg(\frac{1}{d_T} \, \text{Tr}\langle \vert \mathcal{M}_1\vert ^2 \rangle  + \frac{2}{d_T}\, \text{Re}\,\text{Tr} \langle \mathcal{M}_2 \mathcal{M}_0^\ast \rangle    \Bigg) \\
     &= \frac{N}{A_\perp} \frac{1}{d_A} C_2(R) C(R) (4\pi\alpha_s)^2  \int \rho (\Delta z)
\int \frac{d^2 \qperp}{(2\pi)^2} \\& \times
\Bigg( |J(p - \qperp)|^2 \frac{1}{\muperp} + 2 |J(p)|^2 \frac{-1}{2 \muperp^4} \Bigg) \\
&= \frac{N}{A_\perp} \frac{1}{d_A} C_2(R) C(R) (4\pi\alpha_s)^2  \int \rho (\Delta z)
\int \frac{d^2 \qperp}{(2\pi)^2} \\& \times
\Bigg( |J(p - \qperp)|^2 - |J(p)|^2 \Bigg) \frac{1}{ \muperp^4 } 
\end{align}

\subsection{$\boldsymbol{(\mathrm{GLV})_{\mathrm{SUB}}}$ calculation \label{SUBunitarity}} 

    In this section we present the calculation of the ${(\mathrm{GLV})_{\mathrm{SUB}}}$ result. Crucially, in the case of the SUB corrections we make sub-Regge kinematics as outlined in \cref{tab:kin}. 

\subsubsection{Kinematics}
    To see the effect of the sub-Regge kinematics consider the propagator for a parton on shell
    \begin{align}
         (p-q)^2 + i \epsilon &= M^2 ,\\
        p^2 - 2pq +q^2 +i \epsilon &= M^2 ,\\
        -2pq + q^2 + i \epsilon &= 0 \label{eq:prop}.
    \end{align}
    
    In the sub-Regge light-cone coordinates one has 
    \begin{align}
         p\cdot q &= -q^2_z - \frac{\sqrt{2}}{2}P^+ q_z - \qperp^2 \label{eq:sub1}, \\
         q^2 &= - q^2_z - \qperp^2 \label{eq:sub2}.
    \end{align}
    
    substituting \cref{eq:sub1,eq:sub2} into \cref{eq:prop}, one has 
    \begin{align}
        q^2_z + \sqrt{2}P^+ q_z + \qperp^2 +i \epsilon &= 0 .
    \end{align}
    
    Applying the quadratic formula one has,
    \begin{align}
        q^{\pm}_z &= - \frac{\sqrt2}{2}P^+ ( 1 \pm \gamma) \pm i \epsilon \equiv \beta^{\pm},
    \end{align}
    where $ \gamma \equiv \sqrt{1-\frac{2 \qperp^2 }{P^{+2}}} $.
    
    If one Taylor expands $\beta^{\pm}$ to leading order one recovers the familiar GLV result $q^+_z = - \sqrt{2} P^+ $ and $q^-_z = -\frac{\sqrt{2}}{2}\frac{\qperp^2}{P^+} $. By defining the $\gamma$-factor one encodes additional sub-Regge kinematical terms for each pole of the propagator. Therefore the propagator factorizes as
    \begin{align}
        (p-q)^2 + i\epsilon = (q_z - \beta^+)(q_z + \beta^+).
    \end{align}

\subsubsection{Single scattering}

Closing the contour in the negative complex plane the  contributing poles  to the integral in \cref{eq:I1} are 
\begin{align}
    q^{(1)}_z &= \beta^- ,\\
    q^{(2)}_z &= -i\muperp.
\end{align}

with residues

\begin{align}
\text{Res}[I_1,q^{(1)}_{z}]
&= \lim_{ q_{z} \rightarrow  q^{(1)}_{z} }
\frac{1}{(p-q)^2 + i \epsilon}
\frac{4 \pi \alpha_s}{\bigl(q^2_z + \muperp^2\bigr)}
\,e^{-iq_z  \Delta z} ( q_z - q^{(1)}_z) \\[6pt]
&= \lim_{ q_{z} \rightarrow  q^{(1)}_{z} }
\frac{1}{\cancel{(q_z - \beta^-)}(q_z - \beta^+)}
\frac{4 \pi \alpha_s}{\bigl(q^2_z + \muperp^2\bigr)}
\,e^{-iq_z  \Delta z}
\times  \cancel{(q_z - \beta^-)} \\[6pt]
&= \frac{4 \pi \alpha_s}
{(\beta^- - \beta^+)(\beta^- - i\muperp)(\beta^- + i \muperp)}
\,e^{-i \beta^-  \Delta z},
\end{align}

and

\begin{align}
\text{Res}[I_1,q^{(2)}_{z}]
&= \lim_{ q_{z} \rightarrow  q^{(2)}_{z} }
\frac{1}{(p-q)^2 + i \epsilon}
\frac{4 \pi \alpha_s}{\bigl(q^2_z + \muperp^2\bigr)}
\,e^{-iq_z  \Delta z} ( q_z - q^{(2)}_z) \\[6pt]
&= \lim_{ q_{z} \rightarrow  q^{(2)}_{z} }
\frac{1}{(q_z - \beta^-)(q_z - \beta^+)}
\frac{4 \pi \alpha_s}{\bigl(\cancel{q_z + i \muperp}\bigr)(q_z - i \muperp)}
\,e^{-iq_z \Delta z}
\\& \times  \cancel{(q_z + i \muperp)} \\[6pt]
&=  \frac{i4 \pi \alpha_s}
{2\muperp(i \muperp + \beta^-)(i \muperp + \beta^+)}
\,e^{- \muperp \Delta z }.
\end{align}

In the case of the pure sub-Regge kinematical correction we apply the large separation distance approximation and thus neglect the path-length for now, $ \text{Res}[I_1,q^{(2)}_{z}] = 0$ since $ e^{- \muperp \Delta z} \rightarrow 0 $ for $ \muperp \Delta z \gg 1 $,
\begin{align}
    I_1 &= \frac{-2 \pi i}{2 \pi }  \sum^2_{j=1} \text{Res}[{I_1, q^{(j)}_{2,z}}] \\
    &= \frac{-2 \pi i}{2 \pi } ( \text{Res}[I_1,q^{(1)}_{z}] + \text{Res}[I_1,q^{(2)}_{z}]) \\
    &=  \frac{-i4 \pi \alpha_s}{(\beta^- - \beta^+)(\beta^- - i\mu_{\perp})(\beta^- + i \mu_{\perp})} e^{-i \beta^- \Delta z}. 
\end{align}

Furthermore, the phase factor simplifies to $e^{-i \beta^- \Delta z} \approx 1$ since $\Delta z \beta^- \ll 1$. Because $\beta^-$ scales as $\mathcal{O}(\qperp^2 / P^+)$, this is completely consistent with the requirement outlined in \cite{Gyulassy:2000er}; specifically, preserving unitarity demands that $\frac{\Delta z \qperp^2}{P^+} \ll 1$, which inherently guarantees $\Delta z \beta^- \ll 1$. Additionally, this exponential factor reduces to unity when multiplied by its complex conjugate. Consequently, the integral simplifies to
\begin{align}
    I_1   
    &= \frac{-i4 \pi \alpha_s}{(\beta^- - \beta^+)(\beta^{-2} + \muperp^2)}.
\end{align}

The single scattering contribution to the broadening distribution is given by
\begin{align}
    \text{Tr}\langle |\mathcal{M}_1|^2 \rangle &= \frac{N}{A_{\perp}} C_2(R) C(R) (4\pi \alpha_s)^2  \int \rho (\Delta z) \int \frac{d^2 \qperp}{(2\pi)^2} |J(p- \qperpvec )|^2  2P^{+2}
     \\& \times \Bigg[ \frac{1}{(\beta^- - \beta^+)^2(\beta^{-2} + \muperp^2)^2 } \Bigg] .
\end{align}

At this point it is important to note that correction terms has shown up in two ways. First thorugh a scaling type correction $P^+ \rightarrow \beta^+$, and $\frac{\qperp^2}{P^+} \rightarrow  \beta^-$, where the inclusion of the $\gamma$-factor encodes sub-Regge kinematical corrections to each pole. The other form is through additional terms such as $ \muperp^2 \rightarrow \muperp^2 + \beta^{-2}$. These additional terms must still be simplified as keeping such additional terms results in a unitarity violation. Thus the result for $\text{Tr}\langle |\mathcal{M}_1|^2 \rangle$ consistent with unitarity is
\begin{align}
    \text{Tr}\langle |\mathcal{M}_1|^2 \rangle &= \frac{N}{A_{\perp}} C_2(R) C(R) (4\pi \alpha_s)^2 \int \rho (\Delta z) \int \frac{d^2 \qperp}{(2\pi)^2} |J(p- \qperpvec )|^2  2P^{+2}
     \\& \times \Bigg[ \frac{1}{\beta^{+2} \muperp^4 } \Bigg] 
\end{align}

\subsubsection{Double scattering}

Recall the first integral we wish to compute
\begin{align}
    I_2 &= \int \frac{dq_{3,z}}{2 \pi} \frac{4 \pi \alpha_s}{q^2_z + \mu^2_1} \frac{1}{(p - q_3)^2 + i \epsilon} e^{-iq_{3,z} \Delta z}.
\end{align}

This integral takes the same form as $I_1$. We introduce
\begin{align}
    \beta^{\pm}_i &= -\frac{\sqrt{2}}{2} P^+ \Big( 1  \pm \gamma_i \Big) \pm i \epsilon ,\\
    \gamma_i &= \sqrt{1 - \frac{2 q_{\perp,i}^2}{P^{+2}}}.
\end{align}

and therefore 
\begin{align}
    I_2 &= \frac{-i4 \pi \alpha_s}{(\beta_3^- - \beta_3^+)(q_{2,z}^2 + \mu_1^2)}.
\end{align}

For $I_3$,
\begin{align}
        I_3 &= \int \frac{dq_{2,z}}{2 \pi} \Bigg( \frac{-i4 \pi \alpha_s}{(\beta_3^- - \beta_3^+)( q_{2,z}^2 + \mu_1^2)}\Bigg) \frac{1}{(p-q_2)^2 + i \epsilon} \frac{4 \pi \alpha_s }{q^2_{2,z} + \mu^2_2} e^{-iq_{2,z} \Delta z'} .
\end{align}

The poles are 
\begin{align}
    q^{(1)}_{2,z} &= - i \mu_2 ,\\
    q^{(2)}_{2,z} &= -  \beta^-_2 ,\\
    q^{(3)}_{2,z} &= -  \beta^-_3  - i \mu_1.
\end{align}

Therefore the residues

\begin{align}
    \text{Res}[I_3,q^{(1)}_{2,z}] &= \lim_{ q_{2,z} \rightarrow  q^{(1)}_{2,z} }  \Bigg( \frac{-i 4 \pi \alpha_s}{(\beta^-_3-\beta^+_3) [(\beta^-_3 - q_{2,z})^2 + \mu^2_1]} e^{-i \beta^-_3  \Delta z} + \\&  \frac{4 \pi \alpha_s }{2\mu_1 (q_{2,z} - i \mu_1 - \beta^-_3)(q_{2,z} - i\mu_1 - \beta^+_3)} e^{-i(-i\mu_1 + q_{2,z}) \Delta z} \Bigg)\\& \times \frac{1}{(q_{2,z} - \beta^-_2)(q_{2,z}-\beta^+_2)} \frac{4 \pi \alpha_s}{(q_{2,z}-i\mu_2)\cancel{(q_{2,z}+i\mu_2)}} e^{-iq_{2,z} \Delta z'} \cancel{( q_{2,z} + i \mu_2) } \\
    &= \Bigg( \frac{-i 4 \pi \alpha_s}{(\beta^-_3-\beta^+_3) [(\beta^-_3 + i\mu_2)^2 + \mu^2_1]} e^{-i \beta^-_3  \Delta z} \\ &+ \frac{-4 \pi \alpha_s }{2\mu_1 (-i\mu_2 - i \mu_1 - \beta^-_3)(-i\mu_2 - i\mu_1 - \beta^+_3)}  e^{-i(-i\mu_1 + -i\mu_2) \Delta z} \Bigg) \\ & \frac{4 \pi \alpha_s}{(-2i\mu_2)(-i\mu_2-\beta^-_2)(-i\mu_2-\beta^+_2)} e^{-i(-i\mu_2) \Delta z'}.
\end{align}

Since we are computing only the sub-Regge kinematical correction we take $e^{- \muperp \Delta z} \rightarrow 0$, $e^{- \muperp \Delta z'} \rightarrow 1 $ and as argued before $e^{-i \beta^- \Delta z} \rightarrow 1$. Hence,
\begin{align}
   \text{Res}[I_3,q^{(1)}_{2,z}] &=  \frac{ (4 \pi \alpha_s)^2}{(\beta^-_3-\beta^+_3) [(\beta^-_3 + i\mu_2)^2 + \mu^2_1](2\mu_2)(i\mu_2+\beta^-_2)(i\mu_2+\beta^+_2)} ,
\end{align}

and
\begin{align}
    \text{Res}[I_3,q^{(2)}_{2,z}] &= \lim_{ q_{2,z} \rightarrow  q^{(2)}_{2,z} } \Bigg(  \frac{-i 4 \pi \alpha_s}{(\beta^-_3-\beta^+_3) [(\beta^-_3 - q_{2,z})^2 + \mu^2_1]} e^{-i \beta^-_3  \Delta z} \\& + \frac{4 \pi \alpha_s }{2\mu_1 (q_{2,z} - i \mu_1 - \beta^-_3)(q_{2,z} - i\mu_1 - \beta^+_3)} e^{-i(-i\mu_1 + q_{2,z}) \Delta z} \Bigg) \\ & \times \frac{1}{\cancel{(q_{2,z} - \beta^-_2)}(q_{2,z}-\beta^+_2)} \frac{4 \pi \alpha_s}{(q_{2,z}-i\mu_2)(q_{2,z}+i\mu_2)} e^{-iq_{2,z} \Delta z'} \\ & \times \cancel{( q_{2,z} - \beta^-_2) } \\
    &= \Bigg(  \frac{-i 4 \pi \alpha_s}{(\beta^-_3-\beta^+_3) [(\beta^-_3 - \beta^-_2)^2 + \mu^2_1]} e^{-i \beta^-_3  \Delta z} \\ &+ \frac{4 \pi \alpha_s }{2\mu_1 (\beta^-_2 - i \mu_1 - \beta^-_3)(\beta^-_2 - i\mu_1 - \beta^+_3)}  e^{-i(-i\mu_1 + \beta^-_2) \Delta z} \Bigg) \\& \times  \frac{4 \pi \alpha_s}{(\beta^-_2 - \beta^+_2 )(\beta^-_2 - i\mu_2)(\beta^-_2  + i \mu_2)} e^{-i(\beta^-_2 ) \Delta z'}.
\end{align}

Applying the same simplifications,
\begin{align}
     \text{Res}[I_3,q^{(2)}_{2,z}] &= \frac{-i(4 \pi \alpha_s)^2}{(\beta^-_3 - \beta^+_3)[(\beta^-_3 - \beta^-_2)^2 + \mu^2_1](\beta^-_2 - \beta^+_2)(\beta^{-2}_2 + \mu^2_2)},
\end{align}

and
\begin{align}
    \text{Res}[I_3,q^{(3)}_{2,z}] &= \lim_{ q_{2,z} \rightarrow  q^{(3)}_{2,z} } \Bigg( \frac{-i 4 \pi \alpha_s}{(\beta^-_3-\beta^+_3) \cancel{(q_{2,z} - \beta^-_3 + i\mu_1)}(q_{2,z}-\beta^-_3 - i\mu_1)} e^{-i \beta^-_3  \Delta z} \\&+\cancelto{0}{ \frac{4 \pi \alpha_s }{2\mu_1 (q_{2,z} - i \mu_1 - \beta^-_3)(q_{2,z} - i\mu_1 - \beta^+_3)}}  e^{-i(-i\mu_1 + q_{2,z}) \Delta z}\Bigg)   \vspace{4.5cm}\\ & \frac{1}{(q_{2,z} - \beta^-_2)(q_{2,z}-\beta^+_2)} \frac{4 \pi \alpha_s}{(q_{2,z}-i\mu_2)(q_{2,z}+i\mu_2)}  e^{-iq_{2,z} \Delta z'} \\ & \times \cancel{( q_{2,z} - \beta^-_3 + i \mu_1) } \\
    &= \frac{(4 \pi \alpha_s)^2}{2\mu_1 (\beta^-_3 - \beta^+_3)(\beta^-_3 - i \mu_1 - \beta^-_2)(\beta^-_3 - i \mu_1 - \beta^+_2)} \\
    &  \times \frac{1}{(\beta^-_3 - i\mu_1 -i\mu_2)(\beta^-_3 - i\mu_1 +i\mu_2)} e^{-i\beta^-_3 \Delta z} e^{-i(\beta^-_3 - i \mu_1) \Delta z'}.
\end{align}

Applying the same simplifications the residue simplifies to,
\begin{align}
   \text{Res}\big[I_3,q^{(3)}_{2,z}\big] = {} & \frac{(4 \pi \alpha_2)^2}{2\mu_1 (\beta^-_3 - \beta^+_3)(\beta^-_3 - i \mu_1 - \beta^-_2)} \notag \\
   & \times \frac{1}{(\beta^-_3 - i \mu_1 - \beta^+_2)(\beta^-_3 - i\mu_1 -i\mu_2)(\beta^-_3 - i\mu_1 +i\mu_2)} \,.
\end{align}

 One can see at this point that the cross terms for the matrix element will be very complicated. The sub-Regge kinematical correction manifests in two ways. Namely through rescaling and additional terms,
\begin{align}
    P^+ \rightarrow \beta^+ ,\\
    \muperp \rightarrow \muperp + \beta^-.
\end{align}

The presence of these additional terms introduces a fundamental structural inconsistency between the first-order amplitude $\mathcal{M}_1$ and the second-order amplitude $\mathcal{M}_2$. Because these additional contributions appear uniquely in $\mathcal{M}_2$, the initial sources fail to factorize into the characteristic shift structure,

\begin{equation}J(\pperpvec - \qperpvec) \mathcal{F}(\qperpvec) - J(\pperpvec)\mathcal{F}(\qperpvec) = \Big( J(\pperpvec - \qperpvec) - J(\pperpvec) \Big) \mathcal{F}(\qperpvec) ,\end{equation}

thereby manifestly violating unitarity.To restore a physically consistent framework, we focus on the rescaling corrections where the longitudinal momentum is modified via $P^+ \rightarrow \beta^+$. By incorporating this kinematic $\gamma$-factor while systematically dropping the non-factorizable sub-leading terms, the structural factorization of the source current is preserved. Consequently, this prescription guarantees that the complex pole residues take the form:

\begin{align}
   \text{Res}[I_3,q^{(1)}_{2,z}] &=  \frac{ (4 \pi \alpha_s)^2}{(-\beta^+_3) (\mu^2_1 - \mu^2_2)(2\mu_2)(i\mu_2)(\beta^+_2)} ,
\end{align}

\begin{align}
     \text{Res}[I_3,q^{(2)}_{2,z}] &= \frac{-i(4 \pi \alpha_s)^2}{( - \beta^+_3)( \mu^2_1)( - \beta^+_2)( \mu^2_2)},
\end{align}

\begin{align}
   \text{Res}[I_3,q^{(3)}_{2,z}] &= \frac{(4 \pi \alpha_2)^2}{2\mu_1 ( - \beta^+_3)( - i \mu_1 )( - \beta^+_2)(\mu^2_2 - \mu^2_1)},
\end{align}

and therefore,
\begin{align}
    I_3  &= -i   \sum^3_{j=1} \text{Res}[{I_3, q^{(j)}_{2,z}}] =  \frac{-(4 \pi \alpha_s)^2}{2 \mu^2_1 \mu^2_2  \beta^+_3 \beta^+_2 }.
\end{align}

Now, if one applies the averaging over the impact parameter, one has
\begin{align}
    I_3 &=  \frac{-(4 \pi \alpha_s)^2}{2 \muperp^4 \beta^{+2} }.
\end{align}

Crucially, one has the factor of $\frac{1}{2}$. The double scattering contribution to the broadening distribution is given by,

\begin{align}
    \text{Tr}\langle \mathcal{M}_2 \mathcal{M}^{\ast}_0 \rangle &=  \frac{N}{A_{\perp}} C_2(R) C(R) (4\pi \alpha_s)^2 \int \rho (\Delta z) \int \frac{d^2\boldsymbol{q}_{\perp}}{(2\pi)^2} |J(p)|^2 2P^{+2}
    \Bigg[ \frac{-1}{2 \beta^{+2} \muperp^4} \Bigg] .
\end{align}

\subsubsection{The momentum broadening distribution}

The broadening distribution as per \cref{eq:distribution} is therefore given by

\begin{align}
     \frac{dN^{(1)}}{d^3 \vec{p}} &=   \Bigg(\frac{1}{d_T} \, \text{Tr}\langle \vert \mathcal{M}_1\vert ^2 \rangle  + \frac{2}{d_T}\, \text{Re}\,\text{Tr} \langle \mathcal{M}_2 \mathcal{M}_0^\ast \rangle    \Bigg) \\
     &= \frac{N}{A_\perp} \frac{1}{d_A} C_2(R) C(R) (4\pi\alpha_s)^2 \int \rho (\Delta z)
\int \frac{d^2 \qperp}{(2\pi)^2} 2P^{+2} \\ & \times
\Bigg( |J(p - \qperp)|^2  \frac{1}{ \beta^{+2} \muperp}  + 2 |J(p)|^2 \frac{-1}{2 \beta^{+2} \muperp^4}  \Bigg) \\
&= \frac{N}{A_\perp} \frac{1}{d_A} C_2(R) C(R) (4\pi\alpha_s)^2 \int \rho (\Delta z)
\int \frac{d^2 \qperp}{(2\pi)^2} 2P^{+2} \\ & \times 
\Bigg( |J(p - \qperp)|^2 - |J(p)|^2 \Bigg) \frac{1}{ \beta^{+2} \muperp^4 }
\end{align}

Simplifying, one has the substitution 
\begin{align}
    \beta^{+2} &= \frac{1}{2} P^{+2} (1+\gamma)^2 .
\end{align}

Therefore the double scattering contribution to the broadening is
\begin{align}
    \frac{dN^{(1)}}{d^3 \vec{p}} &=   \frac{N}{A_{\perp}} \frac{1}{d_A} C_2(R) C(R)  (4\pi \alpha_s)^2 \int \rho (\Delta z) \int \frac{d^2 \qperp}{(2 \pi)^2} \\ & \times \Bigg(|J(p-\qperp)|^2 - |J(p)|^2 \bigg) \Bigg( \frac{4}{ (1+\gamma)^2} \frac{1}{ \muperp^4}\Bigg)
\end{align}

\subsection{$\boldsymbol{(\mathrm{GLV + APL})}$ calculation}

To perform the APL calculation we use the standard Regge kinematics as in the GLV, \cref{tab:kin}. However, we relax the large system size approximation. Meaning the residues in \cref{I1:zero} and \cref{I2:zero} are no longer 0. 

\subsubsection{Single scattering}

The single scattering residue with the non-zero APL residues

\begin{align}
    \text{Res}[I_1,q_z] &= \sum^2_{j=1} \text{Res}[I_1,q^{(j)}_z] = \frac{4 \pi \alpha_s}{\sqrt{2} P^+ \muperp^2}\Bigg(1-\frac{1}{2}e^{-\muperp \Delta z} \Bigg).
\end{align}

Therefore the integral in \cref{eq:I1} is given by
\begin{align}
    I_1 =  \frac{-i(4 \pi \alpha_s)}{\sqrt{2} P^+ \muperp^2} \Bigg(1-\frac{1}{2}e^{-\muperp \Delta z} \Bigg).
\end{align}

The single scattering contribution is given by
\begin{align}
    \text{Tr}\langle |\mathcal{M}_1|^2 \rangle &= \frac{N}{A_\perp} \frac{1}{d_A} C_2(R) C(R) (4\pi\alpha_s)^2  \int \rho (\Delta z) \int \frac{d^2 \qperp}{(2\pi)^2} \frac{1}{ \muperp^4 }\Bigg(1-\frac{1}{2}e^{-\muperp \Delta z} \Bigg)^2.
\end{align}

\subsubsection{Double scattering }

The residues are once again the same as the GLV case, but with \cref{I2:zero} no longer 0. 

\begin{align}
    \text{Res}[I_2,q_{2,z}] &= \sum^2_{j=1} \text{Res}[I_2,q^{(j)}_{2,z}]\\
    &= \frac{4 \pi \alpha_s}{ \sqrt{2} P^+ (q_{2,z}-i\mu_1)}\Bigg(\frac{1}{q_{2,z}+i\mu_1}-\frac{e^{-i(q_{2,z}-i\mu_1)\Delta z}}{2i \mu_1} \Bigg)
\end{align}

\begin{align}
    I_2 &= \frac{-2\pi i}{2 \pi} \text{Res}[I_2,q_{2,z}]\\
    &= \frac{-i(4 \pi \alpha_s)}{ \sqrt{2} P^+ (q_{2,z}-i\mu_1)}\Bigg(\frac{1}{q_{2,z}+i\mu_1}-\frac{e^{-i(q_{2,z}-i\mu_1)\Delta z}}{2i \mu_1} \Bigg)
\end{align}

Moving to the $I_3$-integral the poles are the same as before as in \cref{pole1,pole2,pole3}.

\begin{align}
    \text{Res}[I_3,q^{(1)}_{2,z}] &= \frac{(4 \pi \alpha_s)}{ \sqrt{2} \mu^2_1 \mu^2_2 P^+} (1 - \frac{1}{2}e^{- \mu_1 \Delta z}),
\end{align}

\begin{align}
    \text{Res}[I_3, q^{(2)}_{2,z}] &= \frac{-(4 \pi \alpha_s)}{2 \sqrt{2}P^+  \mu^2_1(\mu^2_2-\mu^2_1)} e^{-\mu_1 \Delta z'},
\end{align}

\begin{align}
    \text{Res}[I_3, q^{(3)}_{2,z}] = \frac{(4 \pi \alpha_s) }{2 \sqrt{2} P^+ \mu^2_2(\mu_2+\mu_1)} \left( \frac{1}{\mu_2 - \mu_1} + \frac{e^{-(\mu_1 + \mu_2) \Delta z}}{2 \mu_1} \right) e^{- \mu_2 \Delta z'}.
\end{align}

Grouping the relevant terms together and applying the contact limit $\Delta z' \rightarrow 0$,
\begin{align}
    I_3 &= \frac{-i(4 \pi \alpha_s)}{ \sqrt{2}  P^+}
\Bigg( \Bigg[ \frac{1}{\mu^2_1 \mu^2_2} - \frac{1}{2 \mu^2_1 (\mu^2_2-\mu^2_1)}  + \frac{1}{2 \mu^2_2 (\mu^2_2-\mu^2_1)} 
  \Bigg] - \frac{1}{2}\frac{1}{\mu^2_1 \mu^2_2} e^{- \mu_1 \Delta z} \\ & + \frac{1}{4 \mu_1 \mu^2_2 ( \mu_2 + \mu_1)} e^{-(\mu_1 + \mu_2) \Delta z}   \Bigg).
\end{align}

The bracket without any exponential term simplifies as,
\begin{align}
    [\cdots] &= \frac{1}{2 \mu^2_1 \mu^2_2}.
\end{align}

Therefore, applying the contact limit (\cref{delta}) one has,
\begin{align}
    I_3 I_2 &= \frac{-(4 \pi \alpha_s)^2}{4 P^{+2} \muperp^4} \Bigg( 1 - \frac{1}{2} e^{- \muperp \Delta z} \Bigg)^2.
\end{align}

The double scattering contribution to the broadening distribution is therefore given by,
\begin{align}
    \text{Tr}\langle \mathcal{M}_2 \mathcal{M}^{\ast}_0 \rangle &= \frac{N}{A_\perp} \frac{1}{d_A} C_2(R) C(R) (4\pi\alpha_s)^2 \int \frac{d^2 \qperp}{(2\pi)^2}
|J(p)|^2 
\frac{-1}{2\muperp^4} \Bigg( 1 - \frac{1}{2} e^{- \muperp \Delta z} \Bigg)^2.
\end{align}

\subsubsection{The momentum broadening distribution}
Computing the momentum broadening distribution as per \cref{eq:distribution}, one has

\begin{align}
     \frac{dN^{(1)}}{d^3 \vec{p}} &=   \Bigg(\frac{1}{d_T} \, \text{Tr}\langle \vert \mathcal{M}_1\vert ^2 \rangle  + \frac{2}{d_T}\, \text{Re}\,\text{Tr} \langle \mathcal{M}_2 \mathcal{M}_0^\ast \rangle    \Bigg) \\
     &= \frac{N}{A_\perp} \frac{1}{d_A} C_2(R) C(R) (4\pi\alpha_s)^2  \int \rho (\Delta z)
\int \frac{d^2 \qperp}{(2\pi)^2} \\ & \times
\Bigg( |J(p - \qperpvec)|^2 \frac{1}{\muperp} \Bigg( 1 - \frac{1}{2} e^{- \muperp \Delta z} \Bigg)^2  + 2 |J(p)|^2 \frac{-1}{2 \muperp^4} \Bigg( 1 - \frac{1}{2} e^{- \muperp \Delta z} \Bigg)^2 \Bigg) \\
&= \frac{N}{A_\perp} \frac{1}{d_A} C_2(R) C(R) (4\pi\alpha_s)^2  \int \rho (\Delta z)
\int \frac{d^2 \qperp}{(2\pi)^2}
\Bigg( |J(p - \qperpvec)|^2 - |J(p)|^2 \Bigg) \\ & \times \frac{1}{ \muperp^4 } \Bigg( 1 - \frac{1}{2} e^{- \muperp \Delta z} \Bigg)^2.
\end{align}

We note that the double scattering result differs from the single scattering result by a factor of $-\frac{1}{2}$. This is crucial for the factorization and to preserve unitarity.

\subsection{$\boldsymbol{(\mathrm{GLV + APL})_{\mathrm{SUB}}}$ calculation}

We wish to look at the combined result of both the sub-Regge kinematical and All-Path-Length results. The full rersidues are already calculated in \cref{SUBunitarity}. Now we shall proceed to keep the additional path length terms $\propto e^{\-\muperp \Delta z}$ with the new scaling correction $\beta^+$. 

\subsubsection{Single Scattering}

The residues are the same as before. We apply $i \muperp \pm \beta^- \rightarrow i \muperp $ which is needed for unitarity. But we preserve the sub-Regge kinematical correction by including the $\gamma$-factor through $\beta^+$. To preserve unitarity it is required that $e^{-i \beta^- \Delta z} \approx 1$ as $\beta^- \ll 1$ and  $\frac{\Delta z \qperp^2}{P^+} \ll 1$. Therefore the residues become 
\begin{align}
\text{Res}[I_1,q^{(1)}_{z}]
&= \frac{-4 \pi \alpha_s}
{ \beta^+\muperp^2},
\end{align}

and
\begin{align}
\text{Res}[I_1,q^{(2)}_{z}]
&=  \frac{4 \pi \alpha_s}
{2 \beta^+ \muperp^2} e^{- \muperp \Delta z }.
\end{align}

Notably, the new residue of $q^{(2)}_z$, which is no longer 0 and introduces the path length factor $e^{-\muperp \Delta z}$
\begin{align}
    I_1 &=  -i \sum  \text{Res}= \frac{i(4 \pi \alpha_s)}{\beta^+ \muperp} \Big( 1 - \frac{1}{2} e^{- \muperp \Delta z} \Big).
\end{align}

Therefore,
\begin{align}
    \text{Tr}\langle |\mathcal{M}_1|^2 \rangle &= \frac{N}{A_{\perp}} C_2(R) C(R) (4 \pi \alpha_s)^2 \int \rho (\Delta z) \int \frac{d^2 \qperp}{(2\pi)^2} |J(p-\qperpvec )|^2  2P^{+2} \\ & \times
    \Bigg[ \frac{1}{\beta^{+2} \muperp^4} \Big( 1 - \frac{1}{2} e^{- \muperp \Delta z} \Big)^2 \Bigg] .
\end{align}

\subsubsection{Double Scattering}

Applying the same idea to the double scattering residues:
\begin{align}
   \text{Res}[I_3,q^{(1)}_{2,z}] 
   &= \Bigg( \frac{1}{\beta^+_3 (\mu^2_1 - \mu^2_2)} + \frac{1}{2 \mu_1 \beta^+_3 (\mu_1 + \mu_2)} e^{-(\mu_1 + \mu_2) \Delta z} \Bigg) \frac{i(4 \pi \alpha_s)^2}{2\beta^+_2 \mu^2_2},
\end{align}

\begin{align}
     \text{Res}[I_3,q^{(2)}_{2,z}] 
     &= \Bigg( \frac{1}{\beta^+_3 \mu^2_1} - \frac{1}{2\beta^+_3 \mu^2_1 } e^{- \mu_1 \Delta z} \Bigg) \frac{-i(4 \pi \alpha_s)^2}{\mu^2_2 \beta^+_2},
\end{align}

\begin{align}
   \text{Res}[I_3,q^{(3)}_{2,z}] &= \frac{i(4 \pi \alpha_s)^2}{2\mu^2_1 \beta^+_2 \beta^+_3 (\mu^2_2 - \mu^2_1)}.
\end{align}

The integral is given by
\begin{align}
    I_3 &=  -i   \sum^3_{j=1} \text{Res}[{I_3, q^{(j)}_{2,z}}].
\end{align}

Adding the constant terms (terms without $ e^{-\muperp \Delta z}  $ factors).
\begin{align}
    \text{Constant} \Bigg( \sum \text{Res} \Bigg) &= \frac{i(4 \pi \alpha_s)^2}{2 \beta^+_2 \beta^+_3 \mu^2_2 (\mu^2_1 - \mu^2_2)} + \frac{-i(4 \pi \alpha_s)^2}{\beta^+_2 \beta^+_3 \mu^2_1 \mu^2_2} + \frac{i(4 \pi \alpha_s)^2}{2 \beta^+_2 \beta^+_3 \mu^2_1 (\mu^2_2 - \mu^2_1) } \\
    &= \frac{-i(4 \pi \alpha_s)^2}{2\beta^+_2 \beta^+_3 \mu^2_1 \mu^2_2}
\end{align}

The full residue is therefore
\begin{align}
    \sum \text{Res} &= \frac{-i(4 \pi \alpha_s)^2}{2\beta^+_2 \beta^+_3 \mu^2_1 \mu^2_2} + \frac{i(4 \pi \alpha_s)^2}{2 \beta^+_2 \beta^+_3 \mu^2_1 \mu^2_2} e^{- \mu_1 \Delta z} + \frac{-i(4 \pi \alpha_s)^2}{4 \beta^+_2 \beta^+_3 \mu_1 \mu^2_2 (\mu_1 + \mu_2)} e^{-(\mu_1 + \mu_2) \Delta z}.
\end{align}

Applying the average over the impact parameter:
\begin{align}
    I_3 &= -i \sum  \text{Res}\\
    &=  \frac{-(4 \pi \alpha_s)^2}{2\beta^{+2}  \muperp^2 } \Bigg( 1 -  \frac{1}{2} e^{-  \muperp \Delta z} \Bigg)^2 .
\end{align}

And 

\begin{align}
    \text{Tr}\langle \mathcal{M}_2 \mathcal{M}^{\ast}_0 \rangle &= \frac{N}{A_\perp} \frac{1}{d_A} C_2(R) C(R) (4\pi\alpha_s)^2 \int \rho (\Delta z) \int \frac{d^2 \qperp}{(2\pi)^2} 2P^{+2}
|J(p)|^2 \\ & \times
\frac{-1}{2 \beta^{+2} \muperp^4} \Bigg( 1 - \frac{1}{2} e^{- \muperp \Delta z} \Bigg)^2.
\end{align}

\subsubsection{The momentum broadening distribution}

Recall the momentum broadening distribution is given by,

\begin{align}
    \frac{dN^{
     (1)}}{d^3 \vec{p}} &= \text{Tr}\langle |\mathcal{M}_1|^2 \rangle + 2 Re \text{Tr}\langle |\mathcal{M}_2 \mathcal{M}^{\ast}_0 | \rangle\\
    &= \frac{N}{A_\perp} \frac{1}{d_A} C_2(R) C(R) (4\pi\alpha_s)^2 \int \rho (\Delta z) \int \frac{d^2 \qperp}{(2\pi)^2} 2P^{+2} \Bigg( |J(p-\qperpvec)|^2 \\ & \times \frac{1}{ \beta^{+2} \muperp^4} \Bigg( 1 - \frac{1}{2} e^{- \muperp \Delta z} \Bigg)^2 + 2 |J(p)|^2 
    \frac{-1}{2 \beta^{+2} \muperp^4} \Bigg( 1 - \frac{1}{2} e^{- \muperp \Delta z} \Bigg)^2 \Bigg) \\
    &=  \frac{N}{A_{\perp}} \frac{1}{d_A} C_2(R) C(R) (4\pi \alpha_s)^2 \int \rho (\Delta z) \int \frac{d^2 \qperp}{(2 \pi)^2} (|J(p-\qperpvec)|^2 - |J(p)|^2) \\ & \times  \frac{4}{ (1+\gamma)^2} \frac{1}{ \muperp^4} \Bigg( 1 -  \frac{1}{2} e^{-  \muperp \Delta z} \Bigg)^2
\end{align}

\section{Formal subtleties }\label{FormalSubltleties}

    In this appendix, we present results addressing two formal subtleties that arise from the inclusion of sub-Regge kinematical corrections.

\subsection{Variation of the source factor with $\qperp$}

    The standard $2 \rightarrow 2$ scattering process in QCD has a cross-section, 
    \begin{equation}
    \frac{d\sigma}{dt} \propto \frac{1}{s^{2}}.
    \end{equation}
    
    For the diagram in \cref{fig:diagrams}, the relevant kinematics give
        \begin{equation}
            s^{2} = (P^{+} + \qperp)^{4}.
        \end{equation}
    
    The scattering amplitude is taken to have the form
    \begin{equation}
    \frac{1}{2(2\pi)^{3}} |J(p)|^{2}
        = f(E)\, \delta^{(2)}(\pperpvec),
    \end{equation}
    where $f(E)$ contains the momentum dependence of the interaction.  
    Therefore,
    \begin{equation}
    f(p) = \frac{1}{s^{2}}
        = \frac{1}{(P^{+} + \qperp)^{4}}.
    \end{equation}
    
    In the Regge approximation, one assumes that the amplitude varies slowly with the transverse momentum transfer $\qperp$. This leads to the replacement $J(p+ \qperp) \approx J(p)$, which is justified when
    $f(E) \approx \frac{1}{P^{+4}}$,
    since the large power-law suppression makes the energy dependence effectively insensitive to small $\qperp$.\\
    
   However, in the present work, we include sub-Regge kinematical corrections for which this factorization simplification holds up to order $\mathcal{O}(1/P^{+2})$. Sub-eikonal corrections originating from the initial source  scale as $\mathcal{O}(1/P^{+4})$. At $\mathcal{O}(1/P^{+4})$, the explicit dependence on the transverse momentum transfer can no longer be neglected, necessitating the retention of the full non-local structure $J(\pperpvec + \qperpvec)$ rather than the asymptotic eikonal approximation $J(\pperp) \propto \delta^{(2)}(\pperpvec)$. Consequently, incorporating sub-Regge kinematical effects does not require source corrections at the accuracy level considered in this work.
\subsection{ The crossed diagram \label{crossed} } 

\begin{figure}[htp]
\centering
\begin{tikzpicture}
  \begin{feynman}
    % 1. Define Vertices
    \vertex (a) [label={[yshift=+0.5cm]$P^+$}];
    \vertex [right=2.0cm of a] (b);
    \vertex [right=2.5cm of b] (c);
    \vertex [right=2.0cm of c] (d);

    % 2. Scattering Centers
    \node [crossed dot, below=1.5cm of b] (b1);
    \node [crossed dot, below=1.5cm of c] (c1);

    % 3. Draw the initial blob
    \draw[pattern=north east lines] (a) circle (0.25);

    % 4. Draw the main fermion line
    \diagram* {
      (a) -- [fermion, edge label=\(p-q_1-q_2\)] (b)
          -- [fermion, edge label=\(p-q_2\)] (c) 
          -- [fermion, edge label=\(p\)] (d),
    };

    % 5. Draw the Crossed Gluons
    % Use edge label instead of momentum for q2
    \draw [gluon] (c1) -- node[pos=0.6, right, xshift=24pt] {\(q_2\)} (b);
    
    % The visual break (white line) for crossing
    \draw[white, line width=5pt] (b1) -- (c); 
    
    % Use edge label instead of momentum for q1
    \draw [gluon] (b1) -- node[pos=0.6, left, xshift=-24pt] {\(q_1\)} (c);

    % 6. Interaction nodes and z-labels
    \draw[fill=white] (b) circle (0.1);
    \draw[fill=white] (c) circle (0.1);
    
    \node[above=0.2cm of b] {\(z_1\)};
    \node[above=0.2cm of c] {\(z_2\)};

  \end{feynman}
\end{tikzpicture}
\caption{Double scattering diagram with crossed gluon exchanges.}
\end{figure}

One subtlety is that of the crossed diagram. The matrix element is given by

\begin{align}
    \mathcal{M}^{\times}_2 &= i e^{ipx_0} \int \frac{d^2 \qperpone}{(2\pi)^2} \frac{dq_{1,z}}{2 \pi} e^{-i \qperpvec_1 \cdot \bb }\frac{d^2 \qperptwo}{(2\pi)^2} e^{-i \qperpvec_2 \cdot \bb }\frac{dq_{2,z}}{2 \pi}
    J(p-q_1-q_2) \times \\
    & \frac{1}{(p-q_1)^2 + i \epsilon} \frac{1}{(p-q_1 -q_2)^2 + i \epsilon} (\sqrt{2}P^+)^2 v(\vec{q_1}) v(\vec{q_2}) e^{-iq_{1,z}(z_2-z_0)} e^{-iq_{2,z}(z_1-z_0)}
\end{align}

where one notes the change in the exponential factor

\begin{align}
    -i[q_{1,z}(z_2-z_0) + q_{2,z}(z_1-z_0)] &= -i[q_{1,z}(z_2-z_0) + q_{2,z}(z_1 - z_2 + z_2 -z_0)] \\
    &= -i[(q_{1,z} + q_{2,z})(z_2-z_0) + q_{2,z}(z_1 - z_2)]
\end{align}

Calculating the first integral, 

\begin{align}
    I^{\times}_2 &= \int \frac{dq_{2,z}}{2 \pi} \frac{1}{(p-q_1 -q_2)^2 + i \epsilon} v(\vec{q_2}) e^{-i(q_{1,z} + q_{2,z})(z_2-z_0))},
\end{align} 

one has,

\begin{align}
    I^{\times}_2 = \frac{-i(4 \pi \alpha_s)}{\sqrt{2} P^+ (q_{1,z} - i \mu_2)(q_{1,z} + i \mu_2)},
\end{align}

which is the same result as the original GLV calculation but with $1 \leftrightarrow 2$. 

In the case of $I^{\times}_3$,

\begin{align}
    I^{\times}_3 &= \frac{-i 4 \pi \alpha_s}{\sqrt{2} P^+}\int \frac{dq_{1,z}}{2 \pi} \frac{1}{ (q_{1,z} - i \mu_2)(q_{1,z} + i \mu_2)} \frac{1}{(p-q_1)^2 + i \epsilon} \frac{4 \pi \alpha_s}{q^2_{1,z} + \mu^2_1} e^{-i(q_{1,z})\Delta z'}, 
\end{align}

one has $ z_2 - z_1 < 0 $ where previously one had $ z_2 - z_1 > 0$ . One therefore closes the contour in the upper half of the complex plane.

The three poles in the upper half of the complex plane are

\begin{align}
    q^{(1)}_{1,z} = i \mu_1 \\
    q^{(2)}_{1,z} = i \mu_2 \\
    q^{(3)}_{1,z} = \sqrt{2}P^+  
\end{align}

Resulting in,

\begin{align}
    \text{Res}[I^{\times}_3, q^{(1)}_{1,z}=i\mu_1]
    &= \frac{ -i (4 \pi \alpha_s)^2}{2 \mu^2_1 (\sqrt{2}P^+)^2(\mu^2_1-\mu^2_2) } e^{\mu_1 (z_2 - z_1)}
\end{align}

\begin{align}
     \text{Res}[I^{\times}_3, q^{(2)}_{1,z}=i\mu_2] 
    &= \frac{ -i(4 \pi \alpha_s)^2 }{2 \mu^2_2 (\sqrt{2} P^+)^2 (\mu^2_2 - \mu^2_1)} e^{\mu_2(z_2-z_1)},
\end{align}

\begin{align}
     \text{Res}[I^{\times}_3, q^{(3)}_{1,z}=\sqrt{2}P^+] 
    &= \frac{ i(4 \pi \alpha_s)^2}{(\sqrt{2}P^+)^6 }  e^{-i(\sqrt{2}P^+)(z_2-z_1)}
\end{align}

The integral thus becomes,

\begin{align}
    I^{\times}_3
    &=   \frac{ -(4 \pi \alpha_s)^2}{2 \mu^2_1 (\sqrt{2}P^+)^2(\mu^2_1-\mu^2_2) } e^{\mu_1 \Delta z'} + \frac{  -(4 \pi \alpha_s)^2 }{2 \mu^2_2 (\sqrt{2} P^+)^2 (\mu^2_2 - \mu^2_1)} e^{\mu_2 \Delta z'} \\ & + \frac{  (4 \pi \alpha_s)^2}{(\sqrt{2}P^+)^6 }  e^{-i(\sqrt{2}P^+)\Delta z'} .
\end{align}

In the contact limit when $ z_2 - z_1 \equiv \Delta z' \rightarrow 0 $, 

\begin{align}
  I^{\times}_3 &= \frac{ (4 \pi \alpha_s)^2}{(\sqrt{2}P^+)^2}\Bigg( \frac{-1}{2 \mu^2_1 \mu^2_2 } + \frac{1}{(\sqrt{2}P^+)^4 }  \Bigg) 
\end{align}

Consequently, the sub-Regge contribution to the crossed diagram enters at order $\mathcal{O}(1/P^{+4})$, which is two orders higher than the power accuracy targeted in this work. More importantly, retaining this correction would cause the double-scattering result to deviate from the single-scattering baseline, signaling a manifest violation of unitarity.

In the widely separated limit, where the longitudinal distance between scattering centers is macroscopic, the term is dominated by the disparity between scales ($\mu_i \ll P^+ \implies e^{P^+} \gg e^{\mu_i}$). Under these conditions, the integration reduces to
\begin{equation}I^{\times}_3 = \frac{(4 \pi \alpha_s)^2}{(\sqrt{2}P^+)^2} \frac{ e^{-i(\sqrt{2}P^+)\Delta z'}}{(\sqrt{2}P^+)^4}.\end{equation}

Crucially, in the high-energy limit ($P^+ \rightarrow \infty$), the highly oscillatory phase $e^{-i\sqrt{2}P^+\Delta z'}$ is heavily suppressed by the severe power-law scaling of the denominator. As a result, this non-factorizable contribution vanishes rapidly as $I^{\times}_3 \sim \mathcal{O}(1/P^{+4})$, rendering the additional correction term completely negligible in the asymptotic Regge limit.

\section*{7 Acknowledgments}

We thank Ofentse Matlhakola, Cole Faraday, Will Horowitz and Fabio Dominguez for productive discussions and their valuable insights. DVDB and IK thank the National Research Foundation, the National Institute for Theoretical and
Computational Sciences (NITheCS), and the SA-CERN collaboration for their generous
financial support during the course of this work. 
This work is supported by the DST/NRF in South Africa under Thuthuka grant number TTK240313208902.

\section*{8 References}

\bibliographystyle{spphys}
\bibliography{mainnewfinal}  

\end{document}